\documentclass{aa}  

\usepackage{graphicx}
\usepackage{txfonts}
\usepackage{textcomp}
\usepackage[T1]{fontenc}
\usepackage{amsmath}	% Advanced maths commands
\usepackage{amssymb}	% Extra maths symbols
\usepackage{upgreek}
\usepackage{hyperref}
\usepackage{breakurl}
\usepackage{soul}

\begin{document} 

   \title{Near-infrared evolution of the equatorial ring of SN 1987A\thanks{Based on observations made with ESO telescopes at the Paranal Observatory under programme IDs 078.D-0322(A), 086.D-0713(D), 088.D-0638(D), 090.D-0645(E), 094.D-0505(D) and 096.D-0882(D).}}

   %\subtitle{}

   \author{T. Kangas
          \inst{1,2,3}%\fnmsep\thanks{}
          \and
          A. Ahola\inst{2}
          \and
          C. Fransson\inst{4}
          \and
          J. Larsson\inst{3}
          \and
          P. Lundqvist\inst{4}
          \and
          S. Mattila\inst{2,5}
          \and
          B. Leibundgut\inst{6}
          }

   \institute{Finnish Centre for Astronomy with ESO (FINCA), FI-20014 University of Turku, Finland
              \email{tjakangas@gmail.com}
        \and
            Tuorla Observatory, Department of Physics and Astronomy, FI-20014 University of Turku, Finland
        \and
             Department of Physics, KTH Royal Institute of Technology, The Oskar Klein Centre, AlbaNova, SE-106 91 Stockholm, Sweden
         \and
             Department of Astronomy, Stockholm University, The Oskar Klein Centre, AlbaNova, SE-106 91 Stockholm, Sweden
         \and
             School of Sciences, European University Cyprus, Diogenes Street, Engomi, 1516 Nicosia, Cyprus
             \and
             European Southern Observatory, Karl-Schwarzschild-Strasse 2, D-85748 Garching, Germany
             }

   \date{Received ,; accepted ,}

% \abstract{}{}{}{}{} 
% 5 {} token are mandatory
 
  \abstract
  % context heading (optional)
  % {} leave it empty if necessary  
   {
   We use adaptive-optics imaging and integral field spectroscopy from the Very Large Telescope, together with images from the \emph{Hubble Space Telescope}, to study the near-infrared (NIR) evolution of the equatorial ring (ER) of SN~1987A. We study the NIR line and continuum flux and morphology over time in order to lay the groundwork for \emph{James Webb Space Telescope} observations of the system. We also study the differences in the interacting ring structure and flux between optical, NIR and other wavelengths, and between line and continuum emission, to constrain the underlying physical processes. Mostly the evolution is similar in the NIR and optical. The morphology of the ER has been skewed toward the west side (with roughly 2/3 of the NIR emission originating there) since around 2010. A steady decline in the ER flux, broadly similar to the MIR and the optical, is ongoing since roughly this time as well. The expansion velocity of the ER hotspots in the NIR is fully consistent with the optical. However, continuum emission forms roughly 70 per cent of the NIR luminosity, and is relatively stronger outside the hotspot-defined extent of the ER than the optical emission or NIR line emission since 2012--2013, suggesting a faster-expanding continuum component. We find that this outer NIR emission can have a significant synchrotron contribution. Even if emission from hot ($\sim$2000~K) dust is dominant within the ER, the mass of this dust must be vanishingly small (a few $\times10^{-12}$~M$_\odot$) compared to the total dust mass in the ER ($\gtrsim10^{-5}$~M$_\odot$) to account for the observed $HKs$ flux. The NIR continuum emission, however, expands slower than the more diffuse 180-K dust emission that dominates in the MIR, indicating a different source, and the same hot dust component cannot account for the $J$-band emission.
   }


   \maketitle
%
%-------------------------------------------------------------------

\section{Introduction}

Over the past three decades and more, supernova (SN) 1987A has proven to be one of the most important transient events ever observed. Its location only $\sim$50 kpc from us \citep{pietrzynski19}, in the Large Magellanic Cloud (LMC), combined with the high spatial resolution of the \emph{Hubble Space Telescope} (\emph{HST}) and adaptive optics (AO) facilities, has allowed us to study this SN and its transition into a SN remnant (SNR) in a uniquely detailed manner. The insights gained from this follow-up concern topics such as nucleosynthesis \citep[e.g.][]{jerkstrand11}, the late-time evolution of the progenitor system and the different power sources of the SN from radioactive $^{56}$Ni and $^{44}$Ti decay to interaction \citep[for reviews, see e.g.][]{mccray16,milfes17}. Previously, SN~1987A has been observed in the near- and mid-infrared (NIR and MIR, respectively) using ground-based telescopes, the \emph{Stratospheric Observatory for Infrared Astronomy} (\emph{SOFIA}) and the \emph{Spitzer Space Telescope} \citep[e.g.][]{bouchet06,dwek10,arendt16,fransson16,larsson16,matsuura22}, with which it has been possible to study dust, molecules and ejecta morphology. With the recent launch of the \emph{James Webb Space Telescope} (\emph{JWST}), it is now possible to increase the detail of the follow-up even further in the infrared \citep[the first \emph{JWST} study of SN~1987A was presented by][]{larsson23}. With that in mind, an analysis of the time evolution of SN~1987A in the NIR is now timely for setting the baseline for \emph{JWST} studies.

At most wavelengths, the evolving remnant of SN~1987A is dominated by the interaction between the SN ejecta and its circumstellar medium (CSM). The CSM has a well-known triple-ring structure, with an inner equatorial ring (ER) and two outer rings in an hourglass-like shape inclined by $\sim40^\circ$ \citep[e.g.][]{tziamtzis11}. The triple-ring CSM is well replicated by a model in which the matter was ejected 20000 yr ago \citep{crotts00} following a merger of a binary star with initial masses of 14--15 and 7--9~M$_\odot$ \citep{menonheger17,urushibata18}, creating the known immediate progenitor, a blue supergiant of $\sim20$~M$_\odot$. The optical and NIR brightness of the ER has been dominated since about the turn of the millennium by hotspots created by the forward shock of the SN crashing into dense clumps in the ER \citep{borkowski97,sonneborn98,lawrence00}. The hotspots emit especially strongly in optical and NIR lines \citep{kjaer07,groning08}. Since about 2009, these hotspots have been fading \citep{fransson15,arendt16,arendt20} as the fastest ejecta have cleared the ER and now interact with matter outside the ring \citep{larsson19,kangas22}, creating new hotspots there. The fading and destruction of the ER will, however, take some time \citep[e.g.][]{orlando19}.

Dust has a large influence on the infrared emission of the system. Cool dust formed in the ejecta \citep[on the order of 0.5 M$_\odot$ of silicate and carbon dust combined, at $\sim$20~K;][]{indebetouw14,matsuura15,cigan19}\footnote{This is the cold dust mass at $\gtrsim$25 years. Only $\lesssim$10$^{-3}$~M$_\odot$ was claimed to have condensed at 2 years post-explosion \citep{ercolano07}.} has been argued to be responsible for most of the far-infrared (FIR) luminosity of the ejecta, which also provides most of its bolometric luminosity \citep{mccray16}. In the ER, cool dust is less important, but warmer dust dominates the MIR flux \citep[while the inner ejecta contributes very little in the MIR;][]{arendt20} -- the spectral energy distribution is consistent with silicate dust at 180~K \citep{bouchet06,dwek10}, with a weaker, hotter component at shorter wavelengths. Amorphous carbon dust at 525~K was deemed consistent with the hotter component by \citet{arendt16}. Later observations have revealed emerging emission between 30 and 70 $\mu$m at 10000~d, indicative of a dust mass of $>3\times10^{-4}$ M$_\odot$ in the ER \citep{matsuura19}, compared to $\sim10^{-5}$ M$_\odot$ at 8000~d, requiring ongoing dust formation in the ER or possibly a contribution from the warmest ejecta dust. Compared to the gas mass of the ER \citep[$\gtrsim6\times10^{-2}$~M$_\odot$;][]{mattila10}, this would imply a dust-to-gas mass ratio evolving from $\lesssim10^{-3}$ (affected by the destruction of pre-existing dust by SN radiation) to $\lesssim10^{-2}$. The dust responsible for the hotter component, on the other hand, is being destroyed faster than the gas cools \citep{arendt16}; either small grains are evaporated by ultraviolet (UV) photons, or large grains are destroyed in grain-grain collisions. The properties of the dust created by SN~1987A can shed light on the importance of SNe as dust producers in galaxies \citep{indebetouw14}. 

Any contribution of even hotter ER dust in the NIR has not been well studied, and other sources also contribute to the NIR emission, such as thermal free-free emission from shocked gas. Much as in the optical \citep[e.g.][]{fransson13}, numerous narrow emission lines can be seen in the ER spectrum \citep[e.g.][]{kjaer07}, dominated by the shocked and unshocked dense clumps that form the hotspots. Broader emission lines originate from collisionally excited ejecta material crossing the reverse shock. Synchrotron emission from material shocked by the ejecta-ER interaction is dominant in the radio \citep[e.g.][]{indebetouw14,zanardo14,cendes18}, and while the FIR-to-MIR flux of SN 1987A is dominated by dust, a synchrotron spectrum continuing to the NIR might contribute to the continuum there, and is allowed by reported FIR upper limits \citep{cigan19}. These components may be disentangled by studying the evolution of the morphology and spectrum of the NIR emission.

In this paper, we study the morphology and flux evolution of the ER at NIR wavelengths. We use image subtraction to compare the morphology not only between epochs, but also between wavelengths. We especially focus on the NIR continuum emission, which we isolate from and compare to the line emission as well. We describe the data we use in Sect. \ref{sec:data}; and our analysis methods and results in Sect. \ref{sec:analysis}. We discuss our findings in Sect. \ref{sec:disco} and finally present our conclusions in Sect. \ref{sec:concl}. All uncertainties in the paper correspond to $1\sigma$ unless otherwise noted.

%--------------------------------------------------------------------
\section{Observations and data reduction}
\label{sec:data}

\subsection{Imaging observations}

We have used NIR imaging data of SN 1987A taken using NAOS-CONICA \citep[NACO;][]{naco1, naco2} on the Very Large Telescope (VLT) at the European Southern Observatory (ESO) from 2006 to 2017. These observations are listed in \autoref{tab:obslog}. The NACO images were taken using AO in the $J$, $H$, and $Ks$ bands. The NACO data from 2006 are from Program 078.D-0322(A) (PI Danziger), obtained through the public ESO Science Archive\footnote{\url{http://archive.eso.org/eso/eso_archive_main.html}}; the data after 2006 are from Programs 086.D-0713(D), 088.D-0638(D), 090.D-0645(E), 094.D-0505(D) and 096.D-0882(D) (PI Fransson). The individual exposures were taken using a dithering (jitter) pattern of random offsets within a 7-arcsec box around SN 1987A  (8-arcsec for the 2006 images). Due to problems with the NACO detector causing correlated noise in two quadrants which resulted in strong stripes in the images, a dithering pattern was selected for the 2017 observations to align the SN with the two better quadrants. Star 2 \citep{nearbystars,star2} was used as a natural guide star for the AO observations. The pixel scale of the NACO images is 0.013\arcsec /pixel and the field of view 14\arcsec $\times$ 14\arcsec. The full width half maximum (FWHM) angular resolution is listed in \autoref{tab:obslog} for each observation, but we note that the point spread function (PSF) is not well modeled by a single Gaussian due to extended ``wings" common to AO observations.

We have also used public pipeline-reduced NIR images from the Near Infrared Camera and Multi-Object Spectrometer (NICMOS) and Wide-Field Camera 3 (WFC3) instruments aboard the \textit{HST} between 1997 and 2011 (GO 7434, 7821, 9428, 10549, 10867, 11653 and 12241; PI Kirshner). The \textit{HST}/NICMOS images were taken using the filters F110W, F160W, and F205W and obtained from the Mikulski Archive for Space Telescopes (MAST) Portal\footnote{\url{https://mast.stsci.edu/portal/Mashup/Clients/Mast/Portal.html}}. The WFC3 images were taken using F110W and F160W. The WFC3 images were drizzled \citep{drizzle} with a pixel scale of 0.05\arcsec /pixel in the final drizzled image, while in NICMOS images the pixel scale is 0.025~\arcsec /pixel in F110W and F160W and 0.05~\arcsec /pixel in F205W. In order to compare the NIR morphology to that in the optical, we have used images at epochs close to the NACO observations, taken with the Advanced Camera for Surveys (ACS) and WFC3 on the \emph{HST}. These images were taken using the F625W filter between 2006 and 2017 (GO 10867, 12241, 13181, 13810, PI Kirshner; and GO 14753, PI Fransson) and have a final pixel scale of 0.025\arcsec /pixel in the drizzled images. The \emph{HST} images used in this study are listed in \autoref{tab:obslog} as well.

\subsection{NACO image reduction}

The NACO images were reduced with a combination of ESO pipelines\footnote{\url{https://www.eso.org/sci/software/pipelines/naco/}} and Image Reduction and Analysis Facility (\texttt{IRAF})\footnote{\texttt{IRAF} (\url{http://iraf.noao.edu/}) is distributed by the National Optical Astronomy Observatory, which is operated by the Association of Universities for Research in Astronomy (AURA) under cooperative agreement with the National Science Foundation.} tasks as described below. In some of the epochs, the images exhibited a faint horizontal striping pattern, which was first removed from each raw image with a custom \texttt{IRAF} script. For this purpose, the script creates a one-dimensional image by taking the median of the pixel values along each image row. The 400 brightest pixels in each row were excluded from the median to reduce the influence of bright stars. The 1D image was then subtracted from the columns of the original image. The stripe removal was skipped for images from 2015 as the final image quality was not improved.

\begin{figure}
\includegraphics[width=\columnwidth]{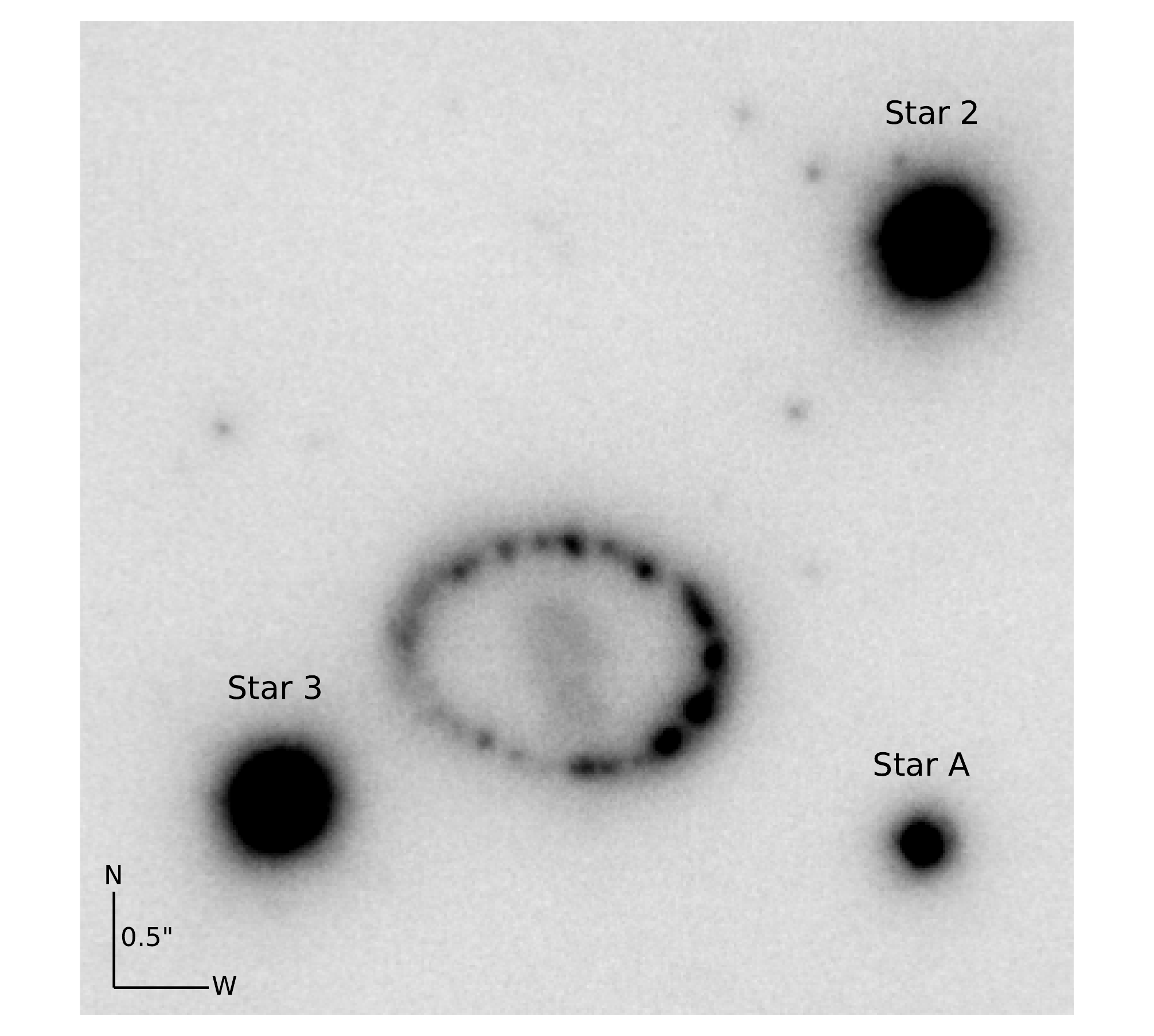}
\caption{A section of the reduced NACO image for the 2010 epoch in the H band, showing the hotspot-dominated ER, the inner SN ejecta in the middle, and nearby bright stars.}
\label{fig:2010hcrop}
\end{figure}

\begin{figure*}
\centering
\includegraphics[width=0.83\linewidth]{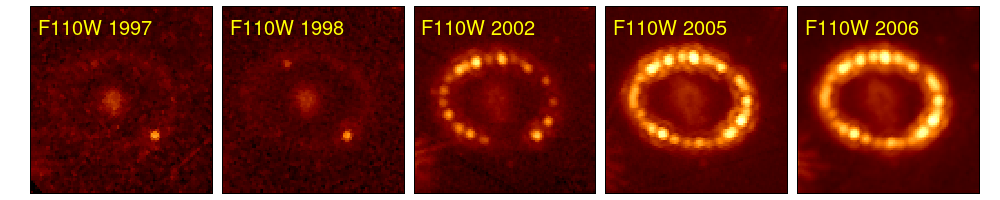}
\includegraphics[width=0.83\linewidth]{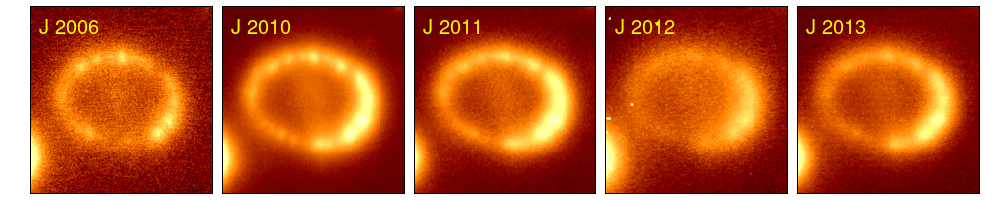}
\includegraphics[width=0.83\linewidth]{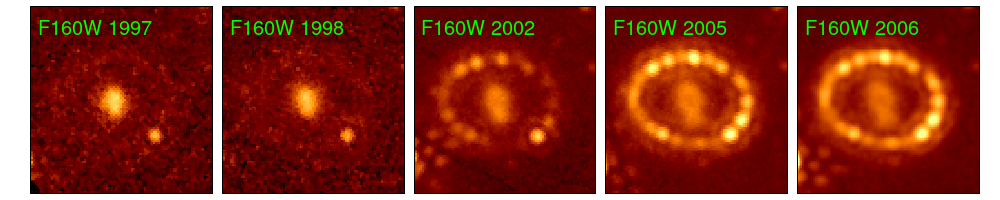}
\includegraphics[width=\linewidth]{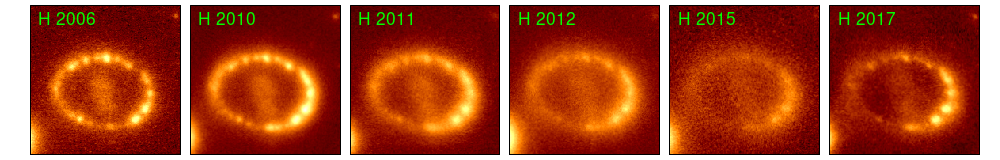}
\includegraphics[width=0.83\linewidth]{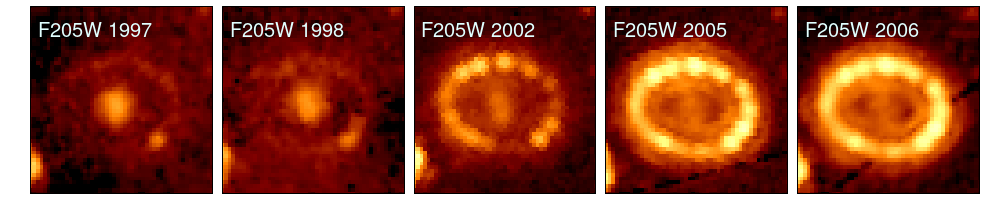}
\includegraphics[width=\linewidth]{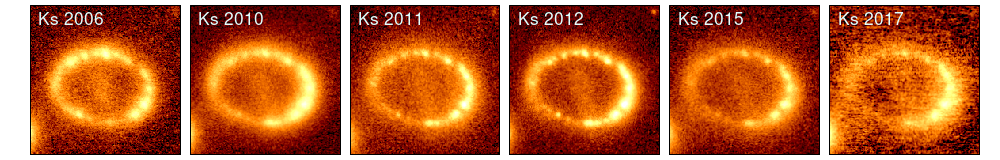}
\caption{The evolution of the morphology of the ER of SN~1987A in NICMOS and NACO images from 1997 to 2017. In each image, north is up and east is to the left. The field of view is $2.5\arcsec$. The color scale in each image is logarithmic.}
\label{fig:all_naco}
\end{figure*}

\begin{figure*}
\centering
\includegraphics[width=0.77\linewidth]{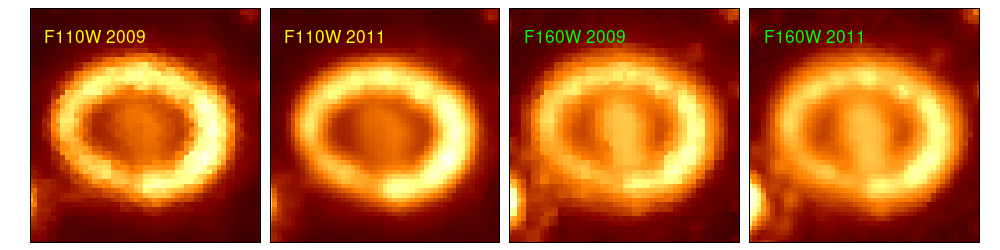}
\caption{WFC3 images from 2009 and 2011. In each image, north is up and east is to the left. The field of view is $2.5\arcsec$. The color scale in each image is logarithmic.}
\label{fig:wfc3}
\end{figure*}

After the stripe removal, the exposures were sky-subtracted in sets of three consecutive exposures to remove time-varying artefacts in the exposures. For $N$ exposures numbered $1,..., N$, the sets were made out of exposures $[1, 2, 3]; [2, 3, 4]; ...; [N-2, N-1, N]$, yielding $N-2$ sets of three exposures. The sky subtraction was carried out one set at a time using the ESO pipeline recipe \texttt{naco\_img\_jitter}, which subtracts a sky frame (the median of the three object frames) from each object frame, aligns the exposures, and stacks them. The recipe yielded one sky-subtracted image for every set of three exposures, $N-2$ in total. The sky-subtracted images were aligned based on the centroid coordinates of a manually selected star in each image. The aligned images were stacked as the median of the images. The sky-subtracted images were visually inspected and images were rejected from the stacking at this step if either individual hotspots or the ring as a whole were not visually discernible. 
Finally, we ran the stripe removal script on the stacked image to remove any bands or stripes left over by the previous steps. \autoref{fig:2010hcrop} shows a section of the final $H$-band image for the 2010 epoch as an example, including the ER, the SN ejecta and nearby bright stars -- Stars 2 and 3 as in \citet[][]{nearbystars, star2}, and a third star, referred to here as Star A. We also show the sequences of ER evolution in all the NACO and NICMOS images we use in this study in \autoref{fig:all_naco}. The WFC3 images have a worse image quality and the evolution is not as apparent from them, but for completeness we show them in \autoref{fig:wfc3}.

\begin{table}
\centering
\caption{Log of imaging observations used in this paper. In NACO images, the seeing FWHM corresponds to the AO-enhanced core of the PSF.}
\label{tab:obslog}
\begin{tabular}{ccccccc}
\hline
Date & Epoch & Filter & Exp. time & FWHM \\
(UT) & (d) & & (s) & (\arcsec) \\
\hline
\textit{HST}/NICMOS & & & & \\
\hline
1997-12-11   & 3944  & F110W  & 256  & 0.10  \\
1997-12-11   & 3944  & F205W  & 2048 & 0.18  \\
1997-12-11   & 3944  & F160W  & 256  & 0.14  \\
1998-07-10   & 4155  & F110W  & 288  & 0.10  \\
1998-07-10   & 4155  & F160W  & 288  & 0.14  \\
1998-07-10   & 4155  & F205W  & 2048 & 0.17  \\
2002-12-31   & 5789  & F110W  & 960  & 0.09  \\
2002-12-31   & 5789  & F160W  & 1600 & 0.14  \\
2002-12-31   & 5789  & F205W  & 1792 & 0.19  \\
2005-11-16   & 6841  & F205W  & 1024 & 0.18  \\
2005-11-16   & 6841  & F110W  & 1408 & 0.10  \\
2005-11-16   & 6841  & F160W  & 1536 & 0.14  \\
2006-12-07   & 7227  & F205W  & 768  & 0.18  \\
2006-12-08   & 7227	 & F110W  & 1536 & 0.10	\\
2006-12-08   & 7227  & F160W  & 1536 & 0.13  \\
\hline
VLT/NACO 	 & 		 & 		  &   	 & \\
\hline
2006-10-10   & 7168  & $Ks$     & 1380 & 0.10  \\
2006-10-13   & 7172  & $J$      & 720  & 0.17  \\
2006-11-11   & 7201  & $H$      & 1200 & 0.10  \\
2010-10-21   & 8641  & $Ks$     & 1620 & 0.10  \\
2010-10-26   & 8645  & $H$      & 2160 & 0.09  \\
2010-10-26   & 8645  & $J$      & 4680 & 0.16  \\
2011-11-05   & 9020  & $H$      & 1560 & 0.13  \\
2011-12-16   & 9061  & $J$      & 2340 & 0.18  \\
2011-12-29   & 9074  & $Ks$     & 1530 & 0.08  \\
2012-12-14   & 9425  & $Ks$     & 2070 & 0.07  \\
2012-12-14   & 9425  & $H$      & 2280 & 0.10  \\
2012-12-14   & 9425  & $J$      & 2340 & 0.22  \\
2013-03-15   & 9516  & $J$      & 2340 & 0.20  \\
2015-02-24   & 10227 & $Ks$     & 4140 & 0.09  \\
2015-02-26   & 10228 & $H$      & 2280 & 0.13  \\
2017-01-14   & 10917 & $Ks$	  & 2070 & 0.09  \\ 
2017-02-02   & 10936 & $H$	  & 3360 & 0.10 \\       
\hline   
\emph{HST}/ACS 	 & 		 & 		  &   	 & \\
\hline
2006-12-06   & 7226  & F625W     & 1200 &  0.06 \\
\hline
\emph{HST}/WFC3 	 & 		 & 		  &   	 & \\
\hline
%2009-12-07  & 8323  & F110W      & 447 & \\ BAD EPOCH!
%2009-12-07  & 8323  & F160W      & 582 & \\ CLEARLY TRACKING PROBLEM IN HST
2009-12-21  & 8337  & F110W      & 447 & 0.19\\
2009-12-21  & 8337  & F160W      & 447 & 0.19\\
2011-01-05   & 8717  & F625W     & 1000 & 0.09  \\
2011-01-05   & 8717  & F110W      & 403 & 0.17 \\
2011-01-05   & 8717  & F160W      & 806 & 0.19 \\
2013-02-06   & 9480  & F625W      & 1200 & 0.09 \\
2015-05-24   & 10317  & F625W      & 1200 & 0.08 \\
2017-08-03   & 11120  & F625W      & 1200 & 0.09 \\
\hline
\end{tabular}
\end{table}

\subsection{Integral field spectroscopy}

We have also studied the ER using integral field unit (IFU) data cubes from the Spectrograph for INtegral Field Observations in the Near Infrared \citep[SINFONI;][]{bonnet03,eisenhauer03} on the VLT. The SINFONI data we have used here, taken in the $J$ band in 2005 and in the $HK$ bands between 2005 and 2017, have been previously published in several papers \citep{kjaer10,fransson16,larsson13,larsson16,larsson19}. We summarize the basic relevant information in \autoref{tab:sinfoni}. The data cubes were reduced (sky-subtracted, wavelength- and flux-calibrated) using ESO pipelines and custom software; for details on the observations and the reduction process, see the aforementioned papers. The systematic uncertainty of SINFONI flux calibration is roughly 10 per cent.

\begin{table}[]
    \centering
    \caption{Log of VLT/SINFONI observations used in this paper. Epochs are averaged in the range of indicated dates weighted by exposure time. Reference numbers correspond to 1: \citet{kjaer10}; 2: \citet{fransson16}; 3: \citet{larsson13}; 4: \citet{larsson16}; 5: \citet{larsson19}.}
    \begin{tabular}{lccccc}
    \hline
         Dates & Epoch & Filter & Exposure & Reference(s) \\
         (UT) & (d) &  & (s) & \\
         \hline
         2005-10-22 & 6826 & $J$ & 4800 & 1, 2, 3, 5\\
          -- 2005-11-18 \\
          2005-10-22 & 6831 & $H$ & 4200 & 1, 2, 3, 5\\
          -- 2005-11-18 \\
          2005-10-31 & 6834 & $K$ & 5400 & 1, 2, 3, 5\\
          -- 2005-11-14 \\
          2007-11-07 & 7604 & $H$ & 4800 & 2, 5\\
          -- 2008-01-09 \\
          2007-11-07 & 7617 & $K$ & 9000 & 2, 5\\
          -- 2008-01-19 \\
          2010-11-05 & 8717 & $H$ & 7800 & 2, 3, 5\\
          -- 2011-01-29 \\
          2010-11-05 & 8697 & $K$ & 6000 & 2, 3, 5\\
          -- 2011-01-02 \\
          2014-10-10 & 10091 & $H$ & 3000 & 2, 4, 5\\
          2014-10-12 & 10120 & $K$ & 7200 & 2, 4, 5\\
          -- 2011-14-01 \\
          2018-01-10 & 11282 & $H$ & 3600 & 5\\
          -- 2018-01-16 \\
          2017-12-09 & 11265 & $K$ & 10800 & 5\\
          -- 2018-01-16 \\
         \hline
    \end{tabular}
    \label{tab:sinfoni}
\end{table}

%--------------------------------------------------------------------
\section{Analysis and results}
\label{sec:analysis}

\subsection{Photometry of the ER and continuum contribution}
\label{sec:phot_cont}

\subsubsection{Measurement and corrections}

\begin{figure}
\centering
\includegraphics[width=0.85\columnwidth]{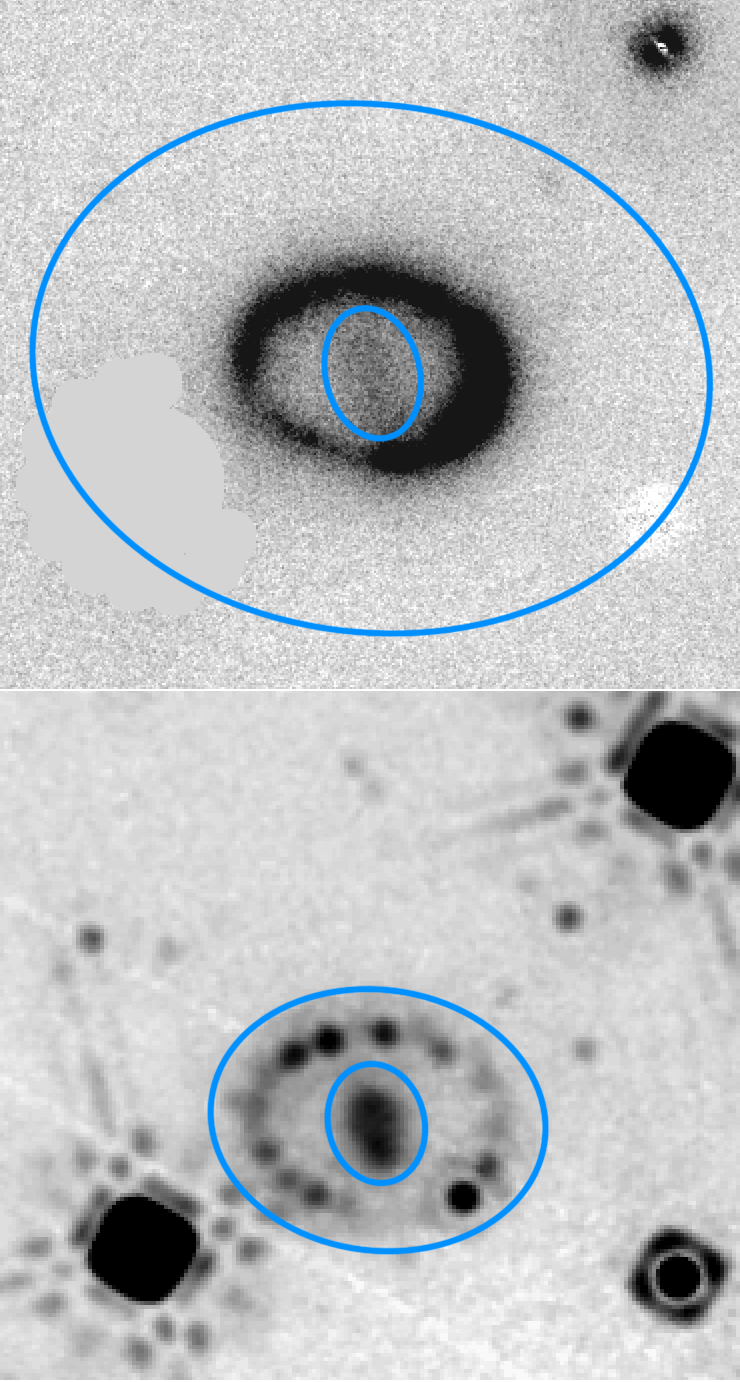}
\caption{The apertures used for measuring the fluxes of the ER and the SN ejecta in the NACO (top) and NICMOS images (bottom). The residuals of two nearby stars from the subtraction of a PSF model are visible in the NACO image. The residual of Star 3 has been manually masked out. North is up and east is to the left.}
\label{fig:ellipses}
\end{figure}

The flux from the ER was measured from images as a difference of fluxes in two elliptical apertures. The larger, outer aperture contained both the ER and the SN ejecta in the middle, and the smaller, inner aperture contained the ejecta in order to remove its contribution. The position angles and sizes for both apertures and the eccentricity for the inner aperture were selected to make them match the shapes of the ER and the ejecta as closely as possible\footnote{Because of the evolving size of the ejecta, the inner aperture was chosen to cover as much of the inner ejecta as possible in 2015 and 2017. The ejecta flux is around 10 per cent of the ER flux in the NACO images, and any systematic effect from ejecta contribution in the ER is small. We also point out that the broad wings of the PSF mean that a small fraction of the ER flux is also within this inner aperture.}. The apertures are shown in \autoref{fig:ellipses}.
For the NACO images, the outer aperture had an eccentricity of 0.63, a semimajor axis of $\sim$2.2$\arcsec$, and a position angle of 7 degrees. The inner aperture had an eccentricity of 0.7, a semimajor axis of $\sim$0.44$\arcsec$, and a position angle of 77 degrees. The eccentricity of the outer aperture corresponds to an axial ratio of 0.78, which was chosen to match the ratio found by \citet{fransson15}. The photometry was carried out in \emph{Starlink} \texttt{GAIA}.
For the NICMOS and WFC3 NIR images, the larger, outer aperture was also an ellipse with an eccentricity of 0.63 and position angle of 7 degrees, but a semimajor axis of $\sim$1.1$\arcsec$ as the PSF does not suffer from AO effects (bottom panel of \autoref{fig:ellipses}). The inner aperture had an eccentricity of 0.6, a position angle of 77 degrees, and a semimajor axis of $\sim$0.4$\arcsec$. 

The fluxes measured from the NACO images were calibrated using known magnitudes of Star 2 (see \autoref{fig:2010hcrop}) from \citet{star2}, assumed to be constant, whereas the \emph{HST} images were flux-calibrated by the pipeline. Errors are not reported in \citet{star2}, and neither are individual NIR measurements of Star 2, but those of the variable Star 3 are. The standard error of the mean\footnote{We use this rather than the standard deviation, as for our NACO photometry we have to assume Star 2 is constant in brightness.} of the Star 3 magnitudes -- after subtracting a linear fit to its decline -- is between 0.015 and 0.033 mag depending on band; as Star 2 is of similar brightness, the precision of its magnitude should also be similar. Considering that the decline of Star 3 may not be exactly linear, we adopt 0.05 mag as the uncertainty for Star 2 (i.e. $J=15.09\pm0.05$, $H=15.08\pm0.05$ and $K=15.06\pm0.05$, Vega). Star 2 might, however, conceivably vary in brightness on a longer timescale than that probed by \citet{star2}. The $JHKs$ magnitudes of the star in the Two-Micron All Sky Survey\footnote{\url{https://irsa.ipac.caltech.edu/Missions/2mass.html}} \citep{2mass} are consistent within $1\sigma$ with the \citet{star2} magnitude. We have also measured the AB magnitudes of Star 2 in the \emph{HST} filters; these are listed in \autoref{tab:star2mag}. With our assumed uncertainty, the AB magnitudes of Star 2 based on \citet{star2} are $J_{\mathrm{AB}} = 16.00\pm0.05$, $H_{\mathrm{AB}} = 16.47\pm0.05$ and $Ks_{\mathrm{AB}} = 16.91\pm0.05$; the measured magnitudes in F110W, F160W and F205W, respectively, stay close to these values from 1997 to 2011, but do show some variation on the order of 0.1 mag over multiple-year timescales (with one outlying epoch in 2006 in F160W), indicating the possibility of similar scatter in NACO $JHKs$ fluxes.

The NACO images have PSFs characteristic to AO, where the narrow, corrected PSF core is surrounded by a fainter, uncorrected halo with broad ``wings". The outer aperture was thus spread far out from the ring to compensate for the PSF. To be able to use a large aperture to measure the ER flux, nearby stars (mainly Star 3 and Star A in \autoref{fig:2010hcrop}) needed to be removed from the images. A PSF fit (using \texttt{SNOoPY}\footnote{\texttt{SNOoPY} (SuperNOva PhotometrY, \url{http://sngroup.oapd.inaf.it/snoopy.html}) is an \texttt{IRAF}-based package designed for performing PSF photometry on SN images.}) was made on the three bright stars closest to the ring. The PSF model was then subtracted from the stars in the image, leaving some residuals at the locations of the removed stars. The residual from Star 3 was removed by replacing the values of the residual pixels with the mean value of the nearby background, measured from an annulus around the residual, using the \texttt{imedit} task in \texttt{IRAF}. 

\begin{table}
\centering
\caption{AB magnitudes of Star 2 at each NICMOS and WFC3 epoch, measured using a 1$\arcsec$ circular aperture.}
\begin{tabular}{ccccc}                                                
\hline
Day & Year & F110W & F160W & F205W \\
\hline
3944 & 1997 & 16.11$\pm$0.01 & 16.64$\pm$0.01 & 17.01$\pm$0.01 \\
4155 & 1998 & 16.20$\pm$0.01 & 16.65$\pm$0.01 & 16.91$\pm$0.01 \\
5789 & 2002 & 16.13$\pm$0.01 & 16.53$\pm$0.01 & 16.97$\pm$0.01 \\
6841 & 2005 & 16.12$\pm$0.01 & 16.55$\pm$0.01 & 17.01$\pm$0.01 \\
7227 & 2006 & 16.08$\pm$0.01 & 16.23$\pm$0.01 & 17.05$\pm$0.01 \\
8337 & 2009 & 16.30$\pm$0.02 & 16.61$\pm$0.03 & - \\
8717 & 2011 & 16.18$\pm$0.02 & 16.53$\pm$0.03 & - \\
\hline      
\end{tabular}
\label{tab:star2mag}  
\end{table}

\begin{table}
\centering
\caption{Continuum fractions (CF) in the ER at each SINFONI epoch.}
\begin{tabular}{ccccc}                                                
\hline
Day & Year & CF($J$) & CF($H$) & CF($Ks$) \\
\hline
6832 & 2005 & 0.65 & 0.76 & 0.66 \\
7612 & 2007 & - & 0.71 & 0.74 \\  
8708 & 2011 & - & 0.73 & 0.75 \\  
10111 & 2014 & - & 0.78 & 0.74 \\  
11269 & 2017 & - & 0.74 & 0.77 \\              
\hline      
\end{tabular}
\label{tab:continuum}  
\end{table}

\begin{figure*}
\centering
\includegraphics[width=0.95\linewidth]{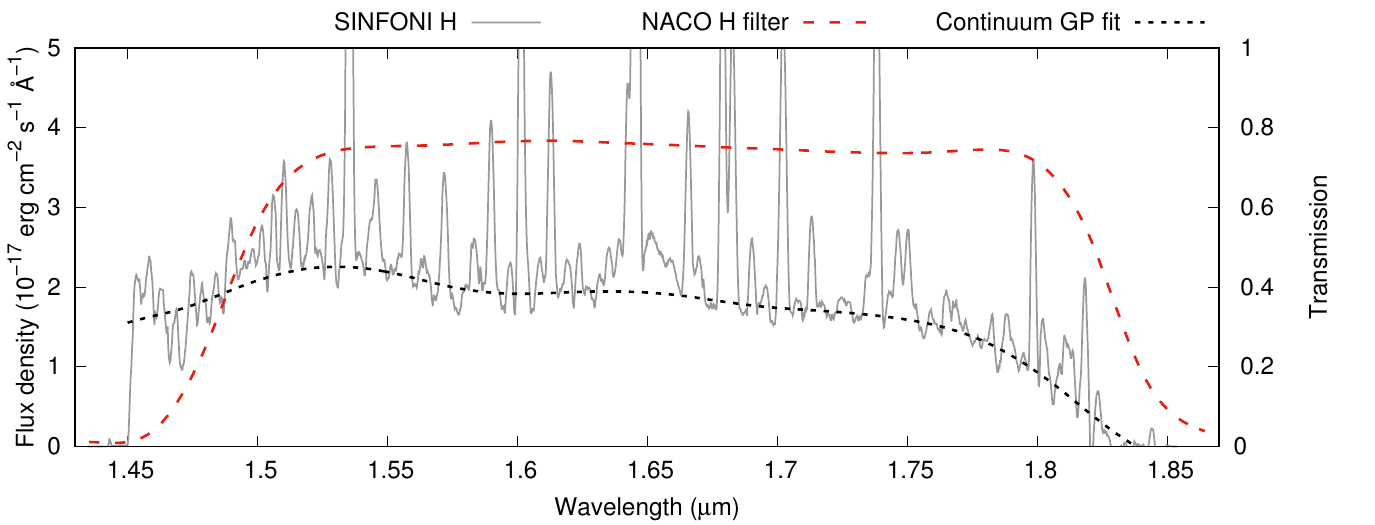}
\caption{Example of a SINFONI $H$-band spectrum from 2017 before (solid grey line) and after line removal and GP fitting (dotted black line). The extent of the NACO $H$ band filter is overplotted (dashed red line). Savitzky-Golay smoothing has been applied to the original spectrum.}
\label{fig:continuum}
\end{figure*}

In addition to the total fluxes from the ER, we have estimated the continuum fluxes. We have determined the fraction of continuum emission contributing to the total flux at each epoch of the SINFONI data, which we have in turn interpolated to correct the NACO fluxes. We extracted the spectrum at each SINFONI epoch and band using \texttt{QFitsView}\footnote{\url{https://www.mpe.mpg.de/~ott/QFitsView/}}, with an aperture covering the entire ER slightly farther out than in the NICMOS photometry (excluding the ejecta region in a similar way) and a sky extraction outside the ER. The semimajor axis was approximately $1.3\arcsec$ and the semiminor axis $1.1\arcsec$, but as the extraction region is set by eye in \texttt{QFitsView}, there is slight variation in the size of the region, on the order of 10 per cent (affecting only the faint emission in the wings of the SINFONI PSF). In each spectrum, we manually removed each clearly detected emission line, down to the surrounding noise level. We then interpolated the remaining continuum-dominated spectrum using a Gaussian process (GP) regression algorithm \citep{rasmussen06}. We used the Python-based {\tt george} package \citep{hodlr}, which implements various different kernel functions. Mat{\'e}rn kernels with $\nu$ parameter of 3/2 or 5/2, common in Gaussian process fitting, were used. We show an example of a SINFONI $H$-band spectrum before and after this process in \autoref{fig:continuum}. The NACO $JHKs$ filters were integrated over the GP-interpolated continuum spectrum and the original, and the ratio between the resulting fluxes is the fraction of continuum flux. The continuum fractions (CF) at each epoch and band are listed in \autoref{tab:continuum}. 

\begin{figure*}
\centering
\includegraphics[width=0.95\linewidth]{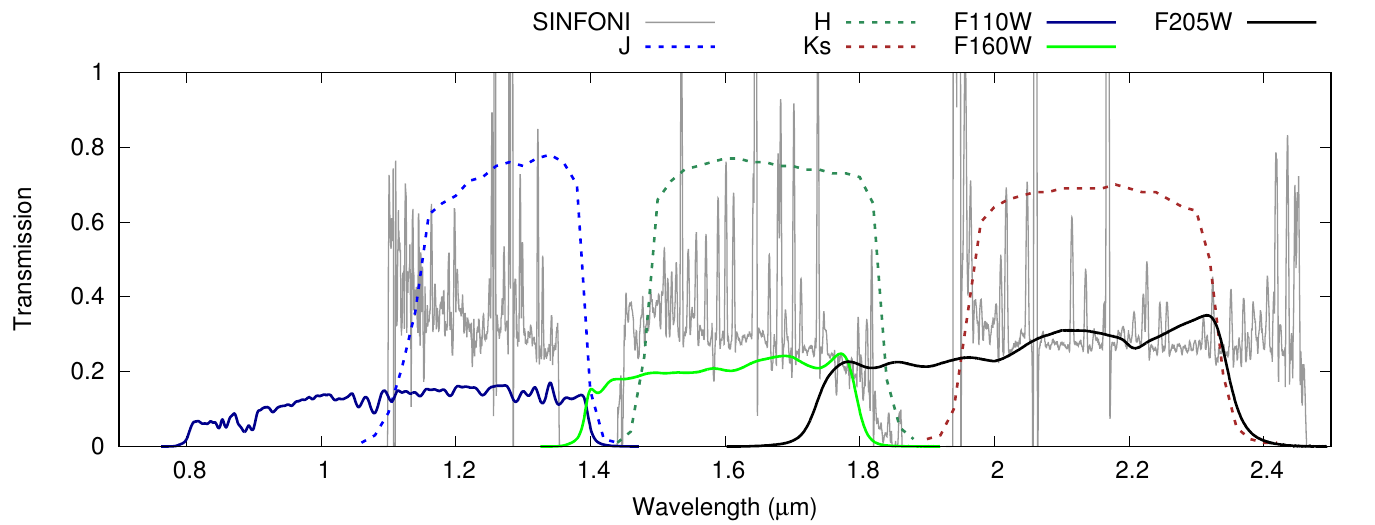}
\caption{Wavelength ranges of NACO and NICMOS filters overlaid on an arbitrarily scaled, Savitzky-Golay smoothed SINFONI $JHK$ spectrum from 2005 \citep{kjaer10}.}
\label{fig:filters}
\end{figure*}

As is clear from \autoref{tab:continuum}, the continuum fraction does not change much between 2005 and 2017, and the linear interpolation of continuum fluxes at the NACO epochs is straightforward. No extrapolation is needed as the NACO epochs all fall within the 6830--11269 day range of the SINFONI observations. For the $J$ band observations, we assume the same fraction as estimated for day 6830 at all epochs; this assumption is unlikely to affect the results much considering the nearly constant CF at the $HKs$ bands. No continuum correction is performed for NICMOS F110W and F205W data, as these filters are considerably wider than the NACO ones, extending outside the SINFONI coverage and thus presumably resulting in a different CF that cannot be determined from SINFONI data in the same way. We illustrate this problem in \autoref{fig:filters}. F160W is closer to $H$, but no SINFONI data exist for interpolating the CF before 6000~d. While the change of the CF over time is slow or nonexistent over the NACO observations, assuming this for the NICMOS epochs as well is unfounded, as the ER flux evolves faster over these epochs.

\begin{figure}
\centering
\includegraphics[width=0.95\columnwidth]{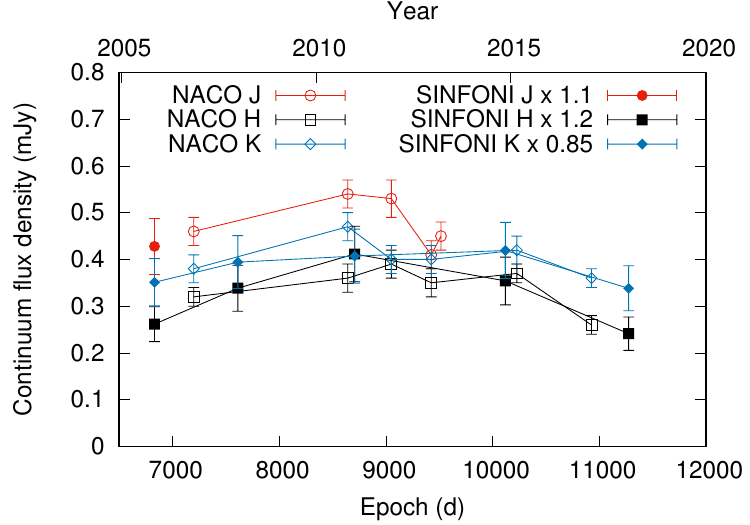}
\caption{Light curves of estimated continuum flux density based on SINFONI and NACO data.}
\label{fig:lc_cont}
\end{figure}

We have also used the SINFONI spectra to determine the total and continuum flux density at those epochs. We have integrated the normalized NACO $JHKs$ filter functions over the GP-fitted SINFONI continuum spectra described above. An assumed 10 per cent uncertainty from the flux calibration and another 10 per cent from extraction and line removal was included in quadrature. The NACO and SINFONI fluxes do not fully agree, as can be seen in \autoref{fig:lc_cont}. A multiplication factor of $\sim$1.1, $\sim$1.2 or $\sim$0.9 must be included in order for the light curves to match over the full span of the observations, depending on the band. The SINFONI data are affected by similar concerns as NACO, namely large PSF wings, which may have an effect in the extraction of the spectrum, but this is not likely to explain the difference completely, and systematic flux calibration effects may be more important. There are also differences between fluxes measured from NICMOS and NACO images around 7200~d -- the NACO flux is lower in $J$ and higher in $H$. This is affected by the different filters, however, so the numbers especially in $J$ and F110W are not directly comparable, and the discrepancy between NACO and NICMOS is similar to that between NACO and SINFONI only in $H$.

\subsubsection{Photometry results}

\begin{figure}
\centering
\includegraphics[width=0.95\columnwidth]{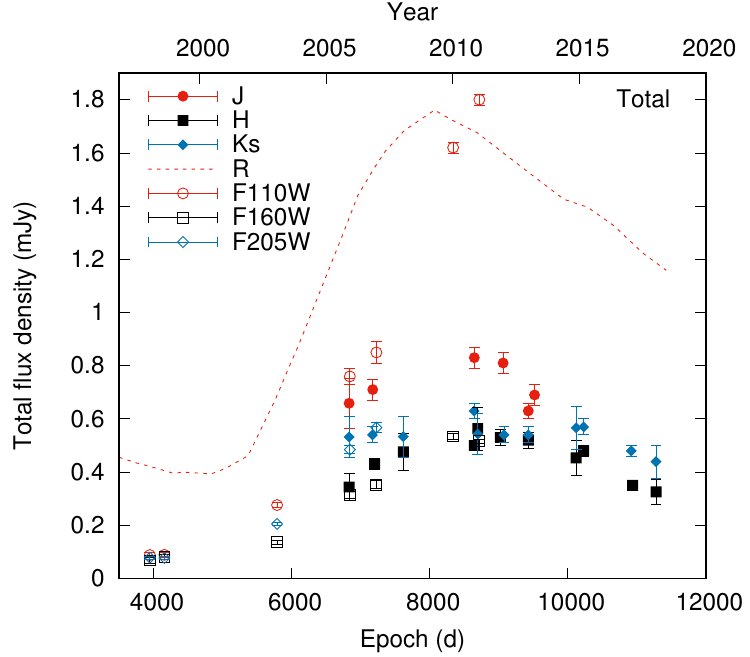}
\includegraphics[width=0.95\columnwidth]{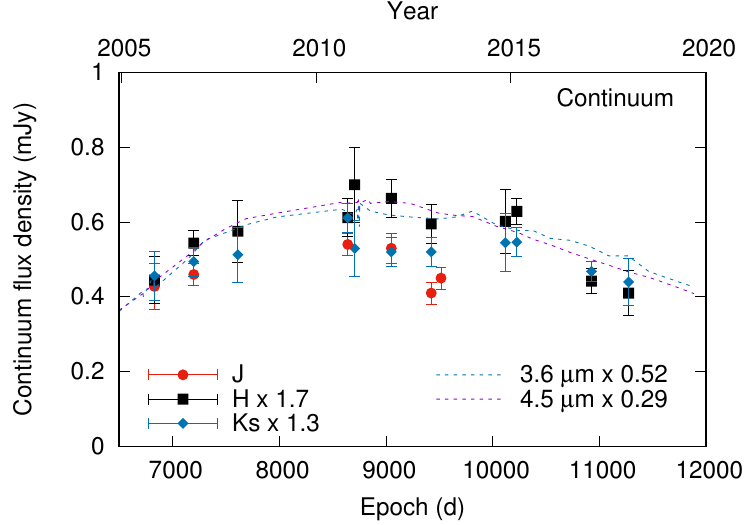}
\caption{Top panel: light curve of the combined ER flux in the NICMOS and NACO bands, compared to the $R$ band \citep{larsson19}. Bottom panel: continuum flux evolution in the NIR, scaled to match at $\sim$7000~d, compared to the MIR light curve, also dominated by continuum \citep{arendt16,arendt20}.}
\label{fig:er_lc}
\end{figure}

We apply the multiplication factors of 1.1 ($J$), 1.2 ($H$) and 0.85 ($Ks$) to fluxes measured from SINFONI, as in \autoref{fig:lc_cont}, to match the light curves as described above. Assuming this correction, we list the ER flux densities in \autoref{tab:ringflux}. We also show the light curve and the evolution of the continuum flux density in the NIR compared to the MIR \citep{arendt16,arendt20} in \autoref{fig:er_lc}. The $JHKs$ light curves peaked between 8000 and 9000~d (i.e. roughly 2010--2012) and declined by about a third in the next 2500~d in $H$ and $Ks$. The optical light curve in $R$ peaked around day 8000 (2010) and in $B$ slightly later; they also declined by roughly a third in the 3000~d after that. The rise toward the peak is also similar in shape. The evolution of the NIR continuum brightness after 7000~d resembles that in the MIR as well, especially in $H$ and $Ks$.

The difference between NACO and \emph{HST} filter functions is apparent from the wildly different light-curve evolution in F110W and $J$ around 2010 ($\sim8500$~d) -- the F110W flux reaches values similar to the $R$ band, a factor of 2 higher than in $J$\footnote{We have checked for problems in the absolute flux calibration of the WFC3/IR images compared to NICMOS or NACO, but the ratio of the ER flux and that of Star 2 is drastically different in F110W and $J$ around 2010 but not in 2006, eliminating this possibility.}. A difference between these filters can be caused by strong emission lines not included in the $J$ filter (i.e. between 8000 and 11000~\AA; see \autoref{fig:filters}); the strongest of these is expected to be He~{\sc i} at 1.083$\mu$m (as the He~{\sc i} 2.058$\mu$m line is the strongest emission line we see in the $K$-band spectra), only a small fraction of which is included in the NACO $J$ filter. We do note that He~{\sc i} lines measured from our SINFONI spectra (see below) do not increase in strength as drastically as the F110W flux does, but the 1.083$\mu$m line was previously noted to be much stronger than expected based on other He lines, owing to different excitation mechanisms \citep{mattila10}. Recent \emph{JWST} observations \citep{larsson23}, with an extremely strong 1.083$\mu$m line from the ER and the RS, corroborate this. Both the soft X-ray \citep{frank16} and H$\alpha$ emission \citep{fransson15} rose by a similar factor as F110W from $\sim$6000 to $\sim$8000~d. We lack the data to say whether the F110W flux kept rising past 9000~d, as did the \textit{hard} X-rays \citep{alp21}. The $H$ and F160W filters are much closer in wavelength coverage, and the light curves in these filters are unsurprisingly much more similar as well.

\begin{table*}
\centering
\caption{ER flux densities ($F$). Numbers in parentheses correspond to the continuum flux density. For NACO $JHKs$ and WFC3 F160W, continuum fluxes have been estimated based on interpolated continuum fractions. In SINFONI fluxes, correction factors of 1.1, 1.2 and 0.85 have been applied in $J$, $H$ and $Ks$ respectively (see text). An uncertainty of 10 per cent has been assumed for SINFONI.}
\begin{tabular}{cccccc}
\hline
Day & Year & Instrument & $F$($J$ / F110W) & $F$($H$ / F160W) & $F$($Ks$ / F205W) \\
 & & & (mJy) & (mJy) & (mJy) \\
\hline
3944 & 1997   &NICMOS& 0.088$\pm$0.009 & 0.068$\pm$0.010   & 0.078$\pm$0.006        \\
4155 & 1998   &NICMOS& 0.089$\pm$0.008 & 0.081$\pm$0.010   & 0.077$\pm$0.006        \\
5789 & 2002    &NICMOS& 0.276$\pm$0.010 & 0.136$\pm$0.008   & 0.205$\pm$0.007        \\
6832 & 2005    & SINFONI & 0.66$\pm$0.10 (0.43$\pm$0.06) & 0.34$\pm$0.05 (0.26$\pm$0.04) & 0.53$\pm$0.08 (0.35$\pm$0.05) \\
6841 & 2005    &NICMOS& 0.76$\pm$0.03   & 0.314$\pm$0.012   & 0.484$\pm$0.015         \\
7168 & 2006    & NACO &       -          &       -            & 0.54$\pm$0.03 (0.38$\pm$0.03)        \\
7172 & 2006    & NACO & 0.71$\pm$0.04 (0.46$\pm$0.03)  &        -           &       -                 \\
7201 & 2006    & NACO &       -          & 0.43$\pm$0.02 (0.32$\pm$0.02)    &        -                \\
7227 & 2006	&NICMOS& 0.85$\pm$0.04   & 0.353$\pm$0.014 & 0.567 $\pm$0.019        \\
7612 & 2007  & SINFONI & - & 0.48$\pm$0.07 (0.34$\pm$0.05) & 0.53$\pm$0.08 (0.39$\pm$0.06) \\
8337 & 2009	&WFC3&  1.62$\pm$0.02  & 0.534$\pm$0.011 (0.390$\pm$0.009) & -        \\
8641 & 2010    & NACO &        -         &       -            & 0.63$\pm$0.03 (0.47$\pm$0.03)        \\
8645 & 2010  	& NACO & 0.83$\pm$0.04 (0.54$\pm$0.03)  & 0.50$\pm$0.02 (0.36$\pm$0.02) &         -               \\
8708 & 2011  & SINFONI & - & 0.56$\pm$0.09 (0.41$\pm$0.06) & 0.54$\pm$0.08 (0.41$\pm$0.06) \\
8717 & 2011	&WFC3&  1.80$\pm$0.02  & 0.516$\pm$0.008 (0.387$\pm$0.006) & -        \\
9020 & 2011    & NACO &        -         & 0.53$\pm$0.03 (0.39$\pm$0.03)  &       -                 \\
9061 & 2011    & NACO & 0.81$\pm$0.04 (0.53$\pm$0.04) &       -            &       -                 \\
9074 & 2011   & NACO &        -         &       -            & 0.54$\pm$0.03 (0.40$\pm$0.03)        \\
9425 & 2012    & NACO & 0.63$\pm$0.03 (0.41$\pm$0.03)  & 0.52$\pm$0.03 (0.39$\pm$0.03)& 0.54$\pm$0.03 (0.40$\pm$0.03)        \\
9516 & 2013    & NACO & 0.69$\pm$0.04 (0.45$\pm$0.03) &        -           &        -                \\
10120 & 2014   & SINFONI & - & 0.45$\pm$0.07 (0.35$\pm$0.05) & 0.57$\pm$0.09 (0.42$\pm$0.06) \\
10227 & 2015   & NACO &       -          &       -            & 0.57$\pm$0.03 (0.42$\pm$0.03)        \\
10228 & 2015  & NACO &        -         & 0.48$\pm$0.02 (0.37$\pm$0.02)   &        -                \\
10917 & 2017  & NACO &       -          &        -           & 0.48$\pm$0.02 (0.36$\pm$0.02)        \\
10936 & 2017  & NACO &        -         & 0.35$\pm$0.02 (0.26$\pm$0.02)  &        -                \\
11275 & 2017  & SINFONI & - & 0.33$\pm$0.05 (0.24$\pm$0.04) & 0.44$\pm$0.07 (0.34$\pm$0.05) \\
\hline
\end{tabular}
\label{tab:ringflux}
\end{table*}

\subsection{Emission line fluxes}

\begin{figure}
\centering
\includegraphics[width=0.99\columnwidth]{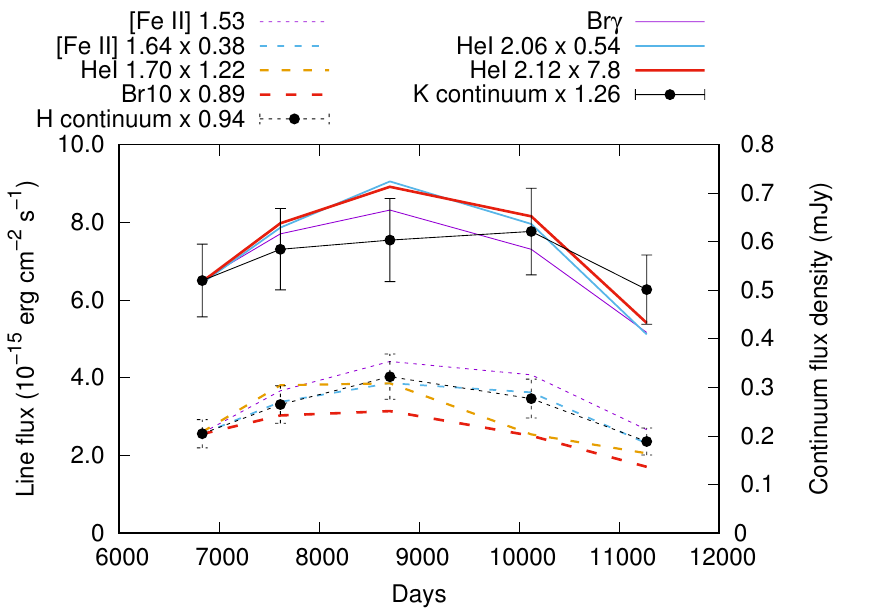}
\caption{Emission line flux evolution over time, compared to the continuum flux density in the $H$ and $K$ bands as measured from the SINFONI spectra. All fluxes are scaled to match in the first SINFONI spectrum.}
\label{fig:line_lc}
\end{figure}

We have measured the fluxes of several narrow emission lines in the SINFONI spectra -- lines originating in both shocked and unshocked matter in the ER (which we cannot distinguish between, but the former becomes dominant over time). As we lack $J$-band spectra after 2005, we concentrate on lines in the other bands. We list the measured line fluxes in \autoref{tab:linefluxes}. In each case, the uncertainty is dominated by the assumed systematic errors we ascribe to the flux calibration and the extraction of the spectrum. We show the evolution of the line fluxes compared to the SINFONI continuum flux density in \autoref{fig:line_lc}. 

The line fluxes mostly evolved similarly to the continuum flux. There are slight variations in evolution between lines, with the largest difference between the [Fe~{\sc ii}] and Br 10 lines in the $H$ band. The flux in all measured lines peaked in the 8708-day spectrum, consistently with the continuum light curve. This is, again, slightly later than in the optical: fluxes of H$\alpha$, [O~{\sc iii}] and [Fe~{\sc xiv}] lines from the shocked ring reported by \citet{fransson15} peaked around 8000~d or, in the case of [O~{\sc iii}], around 7000~d. The lines from \textit{unshocked} matter were disentangled from their high-resolution spectra, but their contribution to the total flux is small and they peaked earlier; thus they do not account for the delay between the NIR and optical peak. 

\begin{table*}
\centering
\caption{Fluxes of relatively strong emission lines in the SINFONI spectra in units of $10^{-15}$~erg~s$^{-1}$~cm$^{-2}$. We include a systematic 10 per cent error from flux calibration and another 10 per cent from extraction in the uncertainties; this is the dominant error for the strong lines. The line fluxes have not been multiplied by a scaling factor.} 
\begin{tabular}{cccccccccc}                                                
\hline
Day & [Fe~{\sc ii}] & [Fe~{\sc ii}] & He~{\sc i} & Br 10 & [Fe~{\sc ii}] & He~{\sc i} & He~{\sc i} & Br$\gamma$ \\
 & 1.533 $\mu$m & 1.644 $\mu$m & 1.701$\mu$m & 1.737 $\mu$m & 2.050 $\mu$m & 2.058 $\mu$m & 2.113 $\mu$m & 2.166 $\mu$m  \\
\hline
6832 & 2.16$\pm$0.33 & 5.72$\pm$0.82 & 1.78$\pm$0.26 & 2.42$\pm$0.36 & 1.16$\pm$0.18 & 14.6$\pm$2.1 & 1.07$\pm$0.18 & 7.9$\pm$1.2 \\
7612 & 3.19$\pm$0.48 & 7.54$\pm$1.08 & 2.54$\pm$0.37 & 2.74$\pm$0.40 & 1.67$\pm$0.24 & 17.8$\pm$2.6 & 1.31$\pm$0.20 & 9.2$\pm$1.4\\  
8708 & 3.86$\pm$0.56 & 8.61$\pm$1.24 & 2.64$\pm$0.39 & 2.98$\pm$0.43 & 1.97$\pm$0.29 & 20.4$\pm$3.0 & 1.47$\pm$0.23 & 10.0$\pm$1.5\\  
10111 & 3.41$\pm$0.49 & 8.01$\pm$1.15 & 1.82$\pm$0.28 & 2.39$\pm$0.36 & 2.23$\pm$0.32 & 17.9$\pm$2.6 & 1.30$\pm$0.20 & 8.5$\pm$1.3\\  
11269 & 2.26$\pm$0.33 & 5.12$\pm$0.73 & 1.39$\pm$0.21 & 1.64$\pm$0.24 & 1.46$\pm$0.21 & 11.7$\pm$1.7 & 0.87$\pm$0.14 & 6.0$\pm$0.9\\              
\hline     
\end{tabular}
\label{tab:linefluxes}  
\end{table*}

\subsection{Expansion velocity of the ER}

We have used the NACO images to determine the size of the ER, and its expansion velocity, in the $Ks$ band where the hotspots can be resolved most consistently. In 2017, on the other hand, the NACO $Ks$ image is visibly of considerably worse quality than in $H$ with fewer measurable hotspots, and we use $H$ instead\footnote{The measured ER sizes are consistent between $H$ and $Ks$ when image quality is similar, in 2006 and 2010, and between NICMOS F160W, $H$ and $Ks$ in 2006.}. We also use F160W on NICMOS, as these images have a better angular resolution than F205W or the WFC3 NIR images (see \autoref{tab:obslog}). The time baseline from our NACO images corresponds to that in \citet{larsson19}; they note that at earlier epochs the size evolution is less linear, and thus epochs before 7000~d are not used in our velocity fit. At each epoch, we have measured the centroids of all visible hotspots and fitted an ellipse to the centroid coordinates. The semimajor axis of the ellipse from this fit has then been converted to physical sizes using a distance of 49.6 kpc to the LMC \citep{pietrzynski19}. 

\begin{figure}
\centering
\includegraphics[width=0.9\columnwidth]{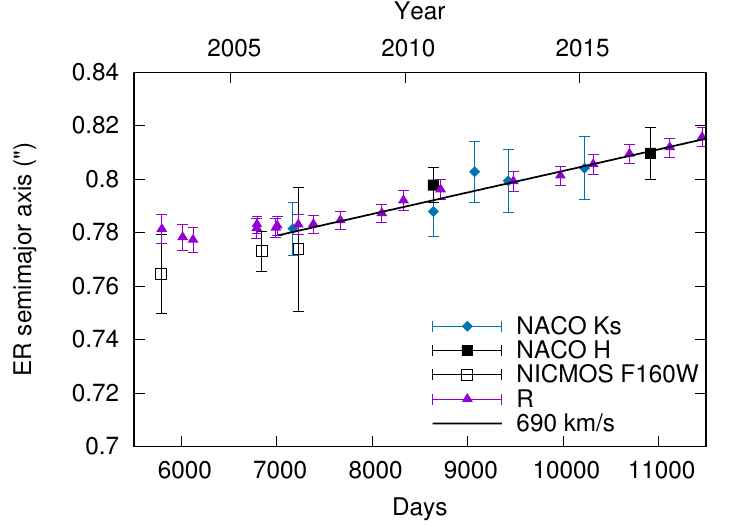}
\caption{ER sizes in the NIR and optical as a function of time \citep[points; optical from][]{larsson19}, and the best expansion velocity fit to the NIR size (line). Only epochs of $>7000$~d are included in the fit. After 7000 d, the ER sizes and expansion velocities determined from optical and NIR hotspots are practically identical, and even before that the sizes are consistent within $\sim1\sigma$.}
\label{fig:expvel}
\end{figure}

We obtain a best-fit expansion velocity of 690$\pm$190 km s$^{-1}$ since 2006 ($\sim$7000~d); we show this fit in \autoref{fig:expvel}. The size evolution appears to have changed around this time in the optical. The measured sizes and the expansion velocity since 2006 are both in excellent agreement with the values measured from optical hotspots by \citet{larsson19}, who obtained a velocity of 680$\pm$50 km s$^{-1}$. This velocity corresponds to the velocity of shocks propagating through dense clumps in the ER. Before 2006 the optical hotspots may be located slightly further out than in the NIR, but any difference between the two is not significant. In the MIR, on the other hand, \citet{matsuura22} measured an expansion velocity of 3920~km~s$^{-1}$, with an initially smaller extent of the MIR ring, but after $\sim$10000~d larger than in the optical. This velocity tracks a more diffuse emission than the optical/NIR hotspots, however.

\subsection{Difference images}

\begin{figure*}
\centering
\includegraphics[width=0.85\linewidth]{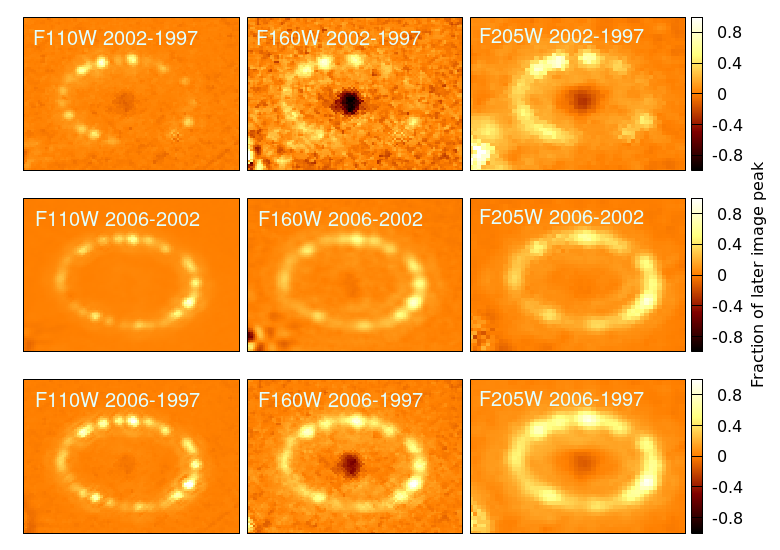}
\caption{Subtractions between selected epochs in the NICMOS F110W, F160W and F205W bands. Each subtraction is scaled to the ER peak counts in the later image used, and the color scale is constant. North is up and east is to the left. Positive pixels indicate brightening between epochs. The field of view is $2.5\arcsec \times 2.1\arcsec$.}
\label{fig:nic_subs}
\end{figure*}

\begin{figure*}
\centering
\includegraphics[width=0.85\linewidth]{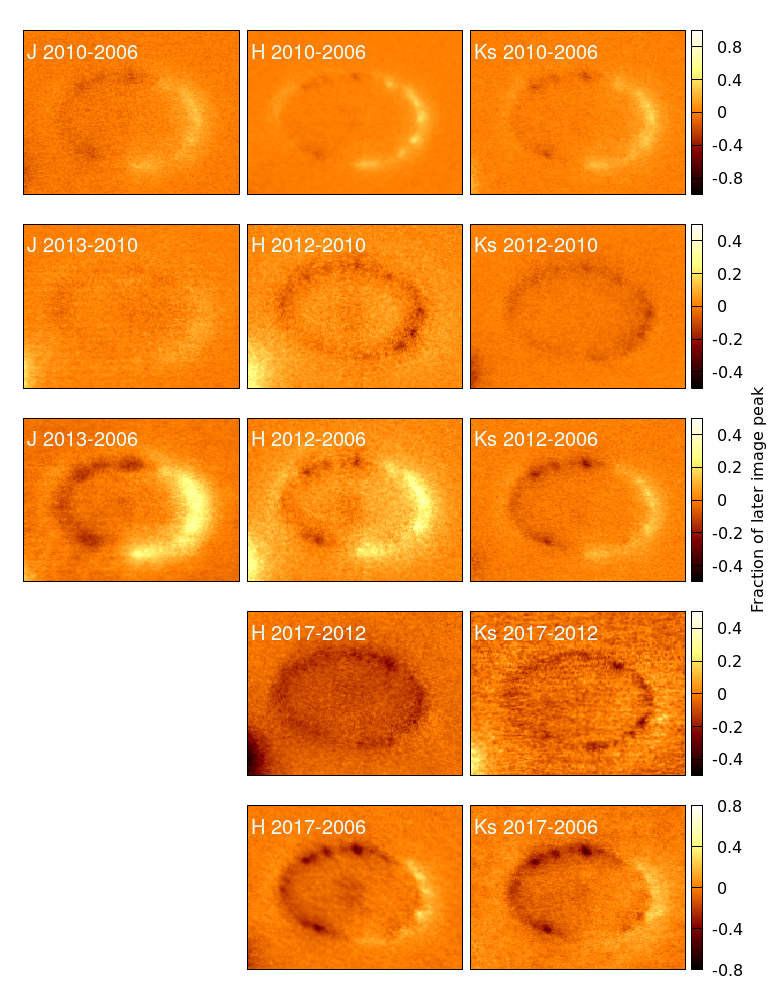}
\caption{Subtractions between selected epochs in the NACO $JHKs$ bands. Each subtraction is scaled to the ER peak counts in the later image used, and the color scale is constant across each row. North is up and east is to the left. Positive pixels indicate brightening between epochs. The field of view is $2.5\arcsec \times 2.1\arcsec$.}
\label{fig:subs}
\end{figure*}

\subsubsection{Evolution of the NIR morphology}

NIR difference images were constructed to visualize the evolution of the SN ejecta and the ER between different epochs. For all of the images, an earlier image was subtracted from a later one in the same band. All of the NACO images were aligned to the 2006 epoch, and the NICMOS images to the 1997 epoch, with the \texttt{IRAF} tasks \texttt{geomap} and \texttt{geotran} prior to the subtraction. The alignment was done using several manually selected bright stars in both images. A general geometry was used which takes into account shift, rotation, skew and distortion between the images. The image subtraction was carried out in a modified version of \texttt{ISIS}\footnote{\url{http://www2.iap.fr/users/alard/package.html}} using the Optimal Image Subtraction method and manual stamp selection. In this method the image with the better seeing is convolved with a kernel derived from comparison with the image with the poorer seeing, yielding two images with similar PSFs and background and intensity levels, allowing for meaningful subtraction of the images \citep{optimal, ois_alard}. Image counts are scaled to match the \textit{later} image in each case. For each subtraction, we used several stars in the field to determine the kernel and the scaling factor, the exact number depending on image quality, i.e. how many stars were visible. We show the resulting subtraction images in Figures \ref{fig:nic_subs} and \ref{fig:subs}; not all epochs are included here, depending on the time between epochs and image quality.

The NICMOS subtractions mainly show the considerable fading of the central ejecta between 1997 and 2002 and the brightening of the ER. The minimum of the optical light curve of the ejecta in fact occurred between these epochs, and it then brightened until at least 2019 \citep[see e.g.][]{larsson19}; however, the central peak of the ejecta still fades over time (and the edges brighten) as the ejecta expands and becomes less centrally peaked (\autoref{fig:all_naco}). The central fading can be seen later as well, including the NACO subtractions, but less clearly, as the \textit{relative} change in ejecta morphology slows down over time.

Between 1997 and 2002, the brightening of the ER occurs mostly on the north and east sides, and between 2002 and 2006, more strongly on the west side. In the 2010$-$2006 NACO subtraction, meanwhile, a stronger change is seen in the ER morphology. The western half of the ER brightens considerably between 2006 and 2010, while the eastern half mostly declines. As is apparent especially in the $H$ band, however, there is also brightening indicating expansion at the easternmost edge. The emission extends roughly equally far from the hotspot peaks on both edges. In the following epochs, the $J$-band image quality is much worse, but in the $H$ and $Ks$ bands a steady decline can be seen all over the ER, except for the ongoing expansion of the ER resulting in a slight brightening toward the advancing edges. A change in the east-west balance as noticeable as that between 2010 and 2006 is not seen afterward, and the ER stays skewed to the west side, which can also be seen in \autoref{fig:all_naco}. The subtractions between each epoch and 2006 show a similar timeline: by 2017, the ER is only brighter than in 2006 at the western and, to a lesser extent, eastern edge, indicating expansion. \citet{fransson15} noted a similar evolution in the optical. 

\subsubsection{Line and continuum maps}

\begin{figure*}
\centering
\includegraphics[width=0.83\linewidth]{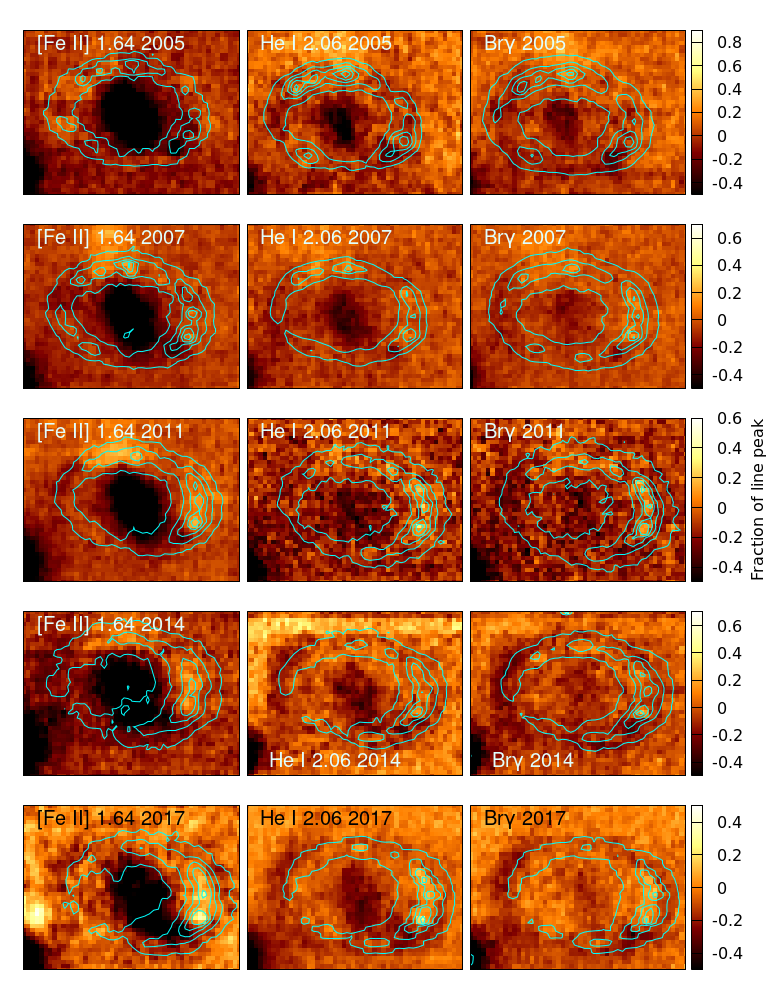}
\caption{Line $-$ continuum difference maps at three strong lines at each epoch, normalized using the average hotspot brightness before subtraction. The contours illustrate the extent of the line emission, and positive fluxes indicate relatively stronger line emission than continuum. The influence of the ejecta and Star 3 is strongest at the relatively weakest ER line of the three, [Fe~{\sc ii}]~$1.64$~$\mu$m. Over time the ER continuum flux moves relatively further out than the line flux, resulting in a clear negative ring slightly outside the line emission in the later epochs. North is up and east is to the left. The color scale is constant across each row. The field of view is $2.5\arcsec \times 2.1\arcsec$.}
\label{fig:linemaps}
\end{figure*}

We have constructed maps of the [Fe~{\sc ii}]~$1.644 \mu$m line, the He~{\sc i}~$2.058 \mu$m line and Br$\gamma$. These were created from the data cubes using \texttt{QFitsView}, subtracting the continuum immediately surrounding the line. In order to study the spatial distribution of the line flux compared to the continuum, we have also extracted maps of the continuum redward of each line. The line maps were constructed by including a range of wavelengths large enough to include the different velocities in different parts of the ER \citep[after correcting for the recession velocity of 287~km$^{-1}$;][]{groning08}; in the case of the [Fe~{\sc ii}] line we included 1.644--1.647~$\mu$m ($\pm$350~km$^{-1}$), in He~{\sc i} 2.0579--2.0640~$\mu$m ($-$350 to 550~km$^{-1}$) and in Br$\gamma$ 2.1657--2.1718~$\mu$m ($-$350 to 550~km$^{-1}$). The continuum map was constructed by, respectively, only including the wavelength ranges 1.652--1.655, 2.0677--2.0751 and 2.1743--2.1804~$\mu$m.

In order to study whether the distribution of the continuum and line emission is the same, in each case we have normalized the maps by matching the average surface brightness of all visible hotspots across the ER between them, then subtracted the continuum from the line map. While the contribution of the continuum to the line map itself has already been subtracted, this comparison clarifies the difference between the spatial distributions -- if the line and continuum originate from the same locations, the subtraction map is practically flat, but any significant positive or negative differences indicate a different spatial distribution. We show the resulting maps in \autoref{fig:linemaps}. The continuum flux gradually becomes stronger relative to the lines slightly outside the ER hotspots, as can be seen via the ring of negative flux in the normalized and subtracted images by 2014, especially in the $K$-band subtractions. Positive hotspots can be seen inside the negative ring in 2014 and 2017, indicating a relatively higher line emission contribution there compared to the region slightly outside the ER. Before this, the extent of the lines is closer to that of the continuum flux, resulting in the negative flux ring / positive hotspot structure being less clear.

\begin{figure*}
\centering
\includegraphics[width=0.9\linewidth]{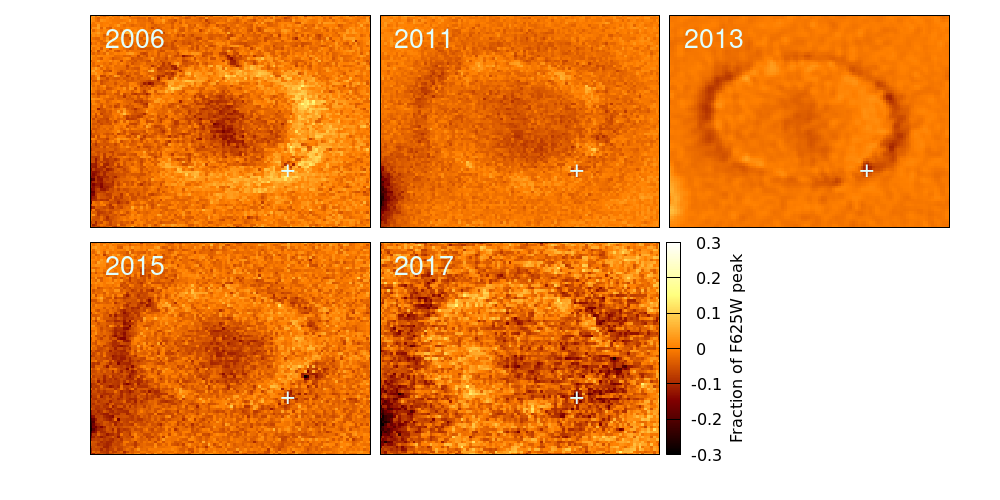}
\caption{F625W-$Ks$ maps at each epoch, illustrating the stronger diffuse emission in $Ks$ since 2013 or so. The images have been convolved to the same seeing using stars, then normalized to the average hotspot brightness in the F625W image before subtraction in each case. Positive pixels indicate relatively stronger F625W- than $Ks$-band emission. The location of a star coinciding with the ER is marked with a $+$ sign. North is up and east is to the left. The field of view is $2.5\arcsec \times 2.1\arcsec$.}
\label{fig:rkmaps}
\end{figure*}

\begin{figure*}
\centering
\includegraphics[width=0.9\linewidth]{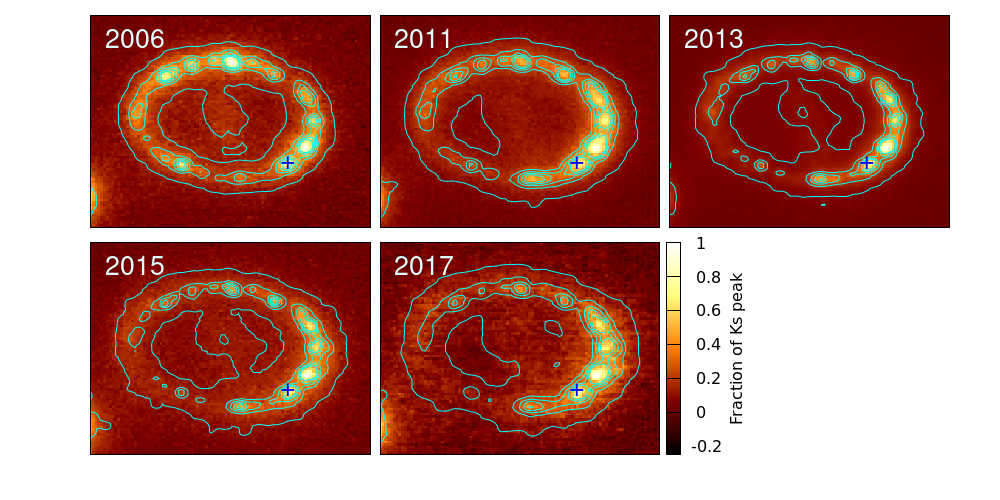}
\caption{F625W contours overlaid on the $Ks$ maps at each epoch after convolving to the same PSF (see text). The location of a star coinciding with the ER is marked with a $+$ sign. North is up and east is to the left. The field of view is $2.5\arcsec \times 2.1\arcsec$.}
\label{fig:rkmaps2}
\end{figure*}

\subsubsection{Differences with optical}

In order to study the differences between optical and NIR emission, we have used the \emph{HST} F625W maps (mostly H$\alpha$ line emission) together with the $Ks$-band maps from NACO. We first performed subpixel alignment, as previously described, between the NIR images and the F625W images closest in time to them. We omit the 2011 NACO epoch here, as the F625W images closest to it in time match the 2010 and 2012 epochs much better and are used in comparisons to them instead (see \autoref{tab:obslog}). We were unable to use \texttt{ISIS} for subtraction, as the colors of the stars in the image can be expected to vary considerably compared to the ER, and they are thus unsuitable for scaling. We have, however, used \texttt{ISIS} to convolve the image with the better seeing in a similar fashion using field stars. We then used the peak counts of the hotspots in the seeing-matched, aligned images to scale the images to each other, then finally subtracted the $Ks$ image from the F625W image at a similar epoch. The resulting maps are shown in \autoref{fig:rkmaps}. We also plot the $Ks$-band images overlaid with contours corresponding to the F625W images in \autoref{fig:rkmaps2}.

A difference is visible in the $Ks$-band outer emission compared to F625W, especially starting in 2013. A ring of $Ks$-band emission is located outside the ER in the subtraction, similar to those in the line-continuum subtractions. The difference is less clear in 2011 and especially 2006; the outer ER emission thus relatively gains strength in the $Ks$ band over time, suggesting a faster expansion. The expansion velocities measured from ellipses fitted to hotspots are consistent (\autoref{fig:expvel}), i.e. the difference is caused by diffuse emission outside the hotspots. The difference is not readily visible from the comparison of the $Ks$ images with the F625W contour maps, however -- only the eastern difference can barely be seen without the subtraction. The eastern ER is less dominated by hotspots than the west, making differences caused by other components more visible.

\subsection{East vs. west fluxes}

\begin{table*}
\centering
\caption{ER flux densities ($F$) in the east (E) and west (W) halves separately at each NACO epoch.}
\begin{tabular}{cccccccc}
\hline
Day & Year & $F_\mathrm{E}$($J$) & $F_\mathrm{W}$($J$) & $F_\mathrm{E}$($H$) & $F_\mathrm{W}$($H$) & $F_\mathrm{E}$($Ks$) & $F_\mathrm{W}$($Ks$) \\
 & & (mJy) & (mJy) & (mJy) & (mJy) & (mJy) & (mJy) \\
\hline
7168 &  2006 & - & - & - & - & 0.21$\pm$0.01 & 0.26$\pm$0.02 \\
7172 &  2006 & 0.22$\pm$0.01 & 0.34$\pm$0.02 & - & - & - & - \\
7201 &  2006 & - & - & 0.15$\pm$0.01 & 0.21$\pm$0.01 & - & -\\
8641 & 2010 & - & - & - & - & 0.19$\pm$0.01 & 0.36$\pm$0.02 \\
8645 & 2010 & 0.30$\pm$0.02 & 0.51$\pm$0.03 & 0.16$\pm$0.01 & 0.30$\pm$0.02 & - & - \\
9020 & 2011 & - & - & 0.19$\pm$0.01 & 0.33$\pm$0.02 & - & -\\
9061 & 2011 & 0.27$\pm$0.02 & 0.50$\pm$0.03 & - & - & - & - \\
9074 & 2011 & - & - & - & - & 0.18$\pm$0.01 & 0.34$\pm$0.02 \\
9425 & 2012 & 0.21$\pm$0.01 & 0.40$\pm$0.02 & 0.16$\pm$0.01 & 0.28$\pm$0.02 & 0.19$\pm$0.01 & 0.34$\pm$0.02 \\
9516 & 2013 & 0.24$\pm$0.02 & 0.43$\pm$0.02 & - & - & - & - \\
10227 & 2015 & - & - & - & - & 0.22$\pm$0.01 & 0.36$\pm$0.02 \\
10228 & 2015 & - & - & 0.19$\pm$0.01 & 0.32$\pm$0.02 & - & -\\
10917 & 2017 & - & - & - & - & 0.16$\pm$0.01 & 0.31$\pm$0.02 \\
10936 & 2017 & - & - & 0.11$\pm$0.01 & 0.24$\pm$0.02 & - & -\\
\hline
\end{tabular}
\label{tab:ringflux_EW}
\end{table*}

\begin{table}
\centering
\caption{East-west ratio at each NACO epoch in each band.}
\begin{tabular}{cccc}
\hline
Year & E/W($J$) & E/W($H$) & E/W($Ks$) \\
 & &  & \\
\hline
2006  &  0.66$\pm$0.05  &  0.73$\pm$0.05  & 0.81$\pm$0.06  \\
2010  &  0.59$\pm$0.04  &  0.53$\pm$0.04  & 0.51$\pm$0.04  \\
2011  &  0.53$\pm$0.04  &  0.58$\pm$0.04  & 0.51$\pm$0.04  \\
2012  &  0.53$\pm$0.04  &  0.59$\pm$0.04  & 0.55$\pm$0.04  \\
2013  &  0.55$\pm$0.04  &  -  & -  \\
2015  & -  &  0.59$\pm$0.04  & 0.60$\pm$0.04  \\
2017  &  -  &  0.47$\pm$0.04  & 0.52$\pm$0.04  \\
\hline
\end{tabular}
\label{tab:ratioEW}
\end{table}

\begin{figure}
\centering
\includegraphics[width=0.95\columnwidth]{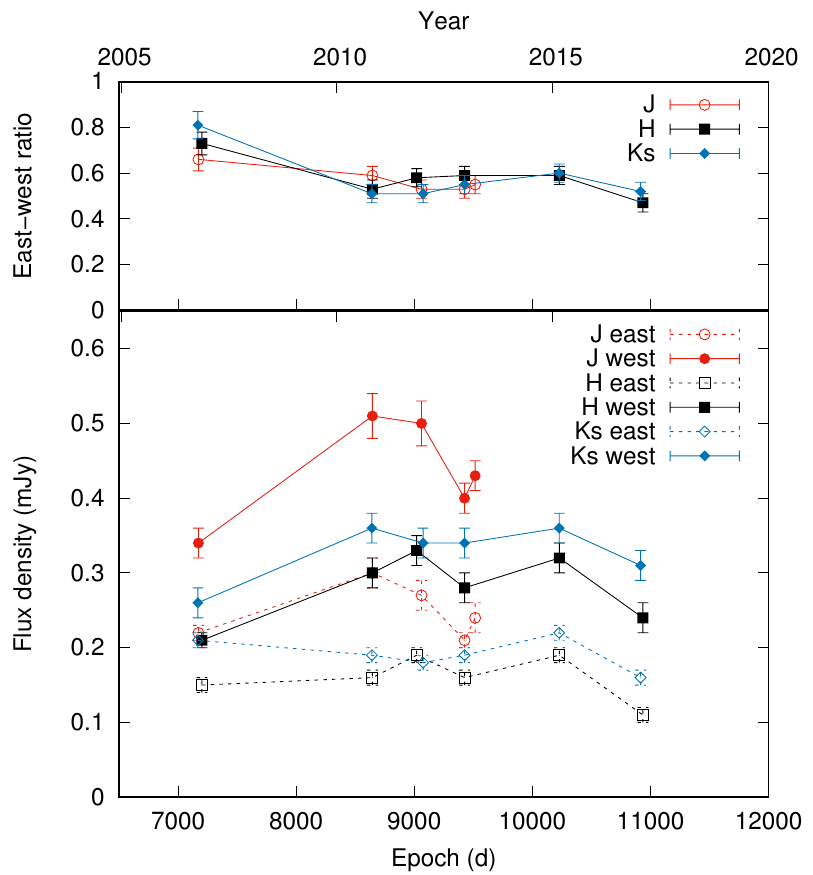}
\caption{Top: evolution of the east-west flux density ratio over time in each band. The clearest change occurs between 2006 and 2010. Bottom: comparison between the flux density in the east (open symbols and dashed lines) and west (solid symbols and lines) halves of the ER. The strongest difference in evolution again occurs between 2006 and 2010, and mainly in the west. Afterward, the evolution is similar in both halves. }
\label{fig:ewratio}
\end{figure}

We have repeated our photometry for the east and west halves of the ER separately in order to quantify the morphological evolution seen in the difference images (\autoref{fig:subs}). We masked out the other half in each NACO image, split at the geometric center of the ER as measured using the flux peaks of the ER at the western and eastern extremes, and repeated the measurement as in Sect. \ref{sec:phot_cont}. The resulting flux densities are listed in \autoref{tab:ringflux_EW}. The east- and west-side light curves and east-west flux density ratios are shown in \autoref{fig:ewratio}; east-west ratios are additionally listed in \autoref{tab:ratioEW}. It is apparent that the east-west ratio changes considerably between 2006 and 2010, similarly to what can be seen in the difference images, but stays fairly constant after 2010. Both halves of the ER decline similarly after the light curve peak.

%--------------------------------------------------------------------
\section{Discussion}
\label{sec:disco}

\subsection{NIR evolution vs. other wavelengths}

Most aspects of the NIR evolution of the ER of SN~1987A follow the optical evolution. The ER light curves in the optical and NIR (and MIR when available) are quite similar, peaking close to the same epoch \citep[$\sim8500$~d i.e. $\sim2010$;][]{fransson15,larsson19} in $J$ and slightly later in $HKs$, and declining at a similar rate after that (\autoref{fig:er_lc}), indicating the same underlying power source, i.e. the ejecta-ER interaction. The evolution of the ER morphology resembles that in the optical as well \citep{fransson15}, with the east-west asymmetry appearing around the time of the light curve peak -- the east side has been declining in the optical since 7000--7500 days ($\sim$2006--2007), and apart from a slight brightening at the eastern edge of the ER due to expansion, our subtractions show the same trend (\autoref{fig:subs}). Furthermore, the expansion of the ER as a whole is also proceeding at the same rate as in the optical when measured from the centroids of hotspots (\autoref{fig:expvel}), i.e. within the hotspots themselves, the locations of optical and NIR emission are practically identical.

The optical east-west ratio in 2006 was $\sim$0.9, but by 2010 it had decreased to $\sim$0.6, and by 2017 below 0.5 \citep{larsson19}. In the NIR, the east-west ratio decreases from 0.7--0.8 in 2006 to 0.5--0.6 since 2010, depending on wavelength. In the soft X-rays \citep{frank16}, the ratio behaves somewhat similarly to its optical and NIR equivalent: $\sim$1.1 in 2006, $\sim$0.8 in 2010 and $\sim$0.6 in 2015. This east-west evolution is also evident in the MIR \citep{matsuura22}, with an increasing fraction of the emission originating in the west over time after roughly 6500~d (2005), albeit going from a ratio of $\sim$3 to $\sim$0.8 instead. This is in stark contrast with the radio \citep[e.g.][]{zanardo18}, millimeter \citep{indebetouw14} and hard X-ray emission \citep{frank16}, where the east side is the brighter one, especially in terms of polarized radio flux \citep[however, according to][the radio asymmetry decreased over time and may have reversed after $\sim$12000~d]{cendes18}. 

However, we see some differences to F625W and MIR emission as well. The diffuse emission \textit{outside} the hotspots in the $Ks$ band is seen to gradually become relatively stronger than in F625W, as shown by the F625W$-Ks$ maps (\autoref{fig:rkmaps}). This only becomes apparent around 2013. As the expansion velocity we measure is based on an elliptical fit to hotspots, this difference has little effect on the velocity measurement. A large fraction of the emission in both bands originates in the hotspots, which results in little visible difference without the subtraction (\autoref{fig:rkmaps2}). The continuum contribution in the $Ks$ band is around 75 per cent in our NACO images, whereas in F625W, the flux is dominated by the narrow and broad components of H$\alpha$ \citep{fransson13}. This contributes to the observed difference if the extended emission in $Ks$ is dominated by continuum, as \autoref{fig:linemaps} indicates.

The expansion velocity of the diffuse ring in the MIR is several times higher than the hotspot expansion velocity in the NIR and optical \citep[$\sim$3900~km~s$^{-1}$ as opposed to $\sim$700~km~s$^{-1}$;][]{matsuura22}, which tracks shocks propagating through dense ER clumps. This is nearly identical to the velocity in the radio \citep{zanardo13}, while the X-rays run the gamut from $\sim$3100~km~s$^{-1}$ in $>2$~keV X-rays \citep{frank16}, through $\sim$1900~km~s$^{-1}$ between 0.5 and 2 keV, to consistent with zero (or even with slow shrinking) in the 0.3--0.8 keV band. The MIR ring was initially smaller than in the NIR before about 10000~d ($\sim$2014), when it caught up to the extent of the hotspots, and larger afterward. However, the more diffuse NIR emission component outside the hotspots, responsible for the seemingly larger extent of the ER compared to the optical, may track this faster expansion. Even so, despite the similar light curve, most of the NIR emission thus seems to originate in a different location than the 180~K dust that dominates the MIR (or the X-ray-emitting shocked gas, except for the softest X-rays). 

In the MIR, the brightness difference is associated with a shift in temperature: the dust on the east side was hotter in 2005 and on the west side in 2017 \citep{matsuura22}, possibly because of a density difference, which could also explain the stronger NIR/optical emission from the interaction on the west side. \citet{larsson23} also found that the extent of the reverse shock is largest on the southeast side, the faintest section of the ER with the fewest visible hotspots in our images, indicating a lower density there.

\subsection{The properties of the NIR continuum}

The continuum consistently forms a majority of the total NIR emission from the ER between 6800 and 11300~d (2005--2017), and possibly earlier as well. We have showed that the spatial extent of the continuum emission is not identical to the line emission. Hotspots in the ER are responsible for much of both the lines and the continuum emission, as is apparent from our images, the lack of evolution in the CF and hence the similarity of the continuum and line emission light curves (see \autoref{fig:line_lc}). However, the continuum is relatively stronger outside the hotspots than the line emission as well \autoref{fig:linemaps}. This, together with the larger CF in the NIR than in the optical as mentioned above, indicates that the difference between F625W and NIR, with NIR emission similarly being stronger beyond the hotspots, is specifically caused by the continuum emission. 

The MIR continuum of the ER is dominated by 180~K silicate dust \citep{dwek10}, collisionally heated by gas which itself is heated by the ejecta-ER interaction and subsequently emits X-rays. Emission from such collisionally heated dust should thus follow the X-ray evolution, but the MIR/X-ray flux ratio (and, similarly, the NIR/X-ray ratio) lessens over time, indicating destruction of dust by sputtering (i.e. dust-gas collisions within shocked gas that return the dust grain material into the gas phase). An additional hotter component is also seen with $T\gtrsim350$~K. \citet{arendt16} fit this component with 525-K amorphous carbon dust, which is being destroyed over time faster than the gas cools -- either by UV photons or by grain-grain collisions, depending on the grain size. These components are not responsible for most of the $JHKs$ emission, however -- nor are they spatially coincident with it. 

\begin{figure}
\centering
\includegraphics[width=0.97\columnwidth]{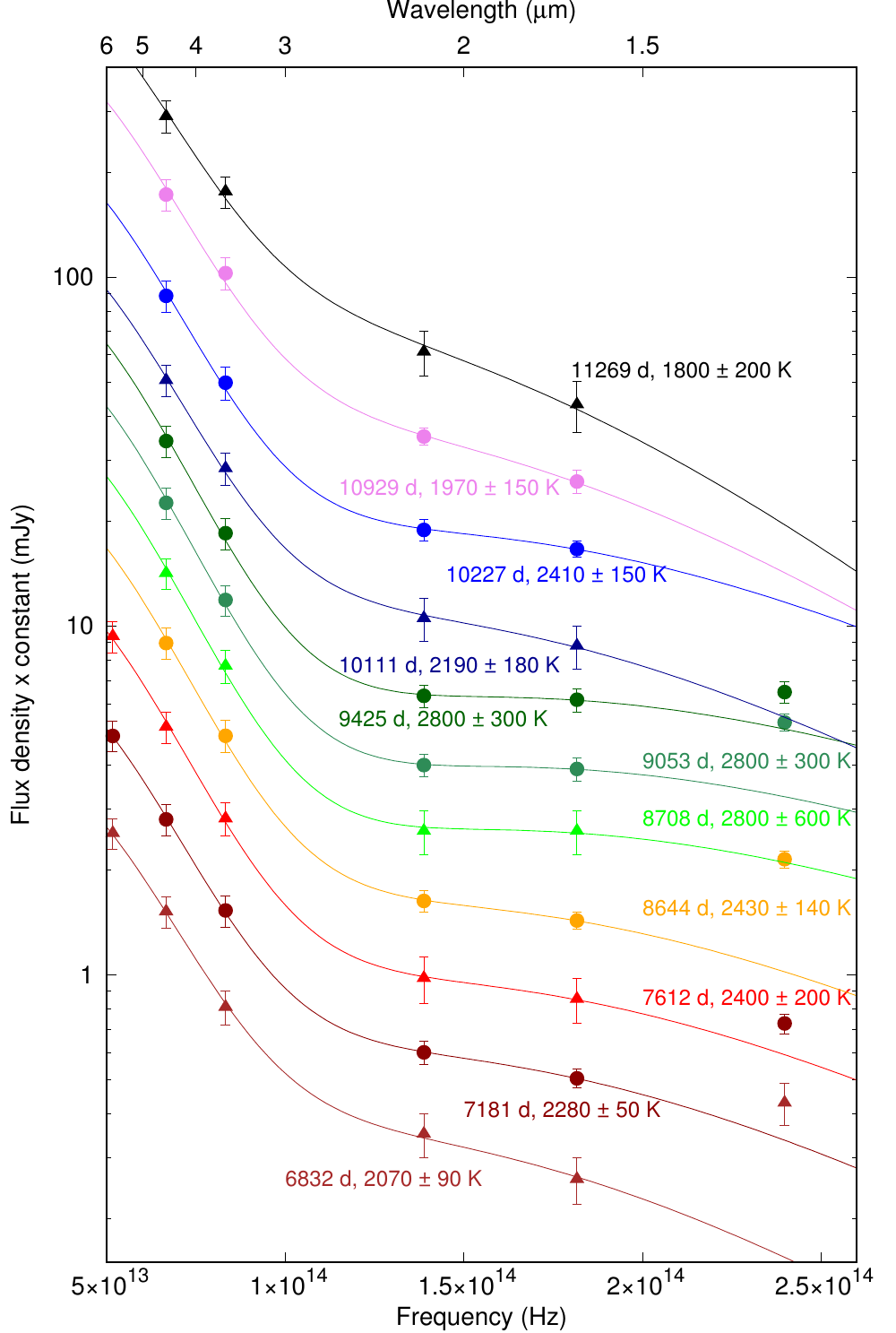}
\caption{Two-component blackbody fits to the MIR and NIR flux densities, with the cooler dust fixed at a temperature of 525 K \citep{arendt16}. The $J$-band points clearly do not fit the same blackbody function as the $HKs$ points in most cases, and are not included in the fits. The SINFONI epochs (with the correction factor included) are plotted with triangles and the NACO epochs with circles. For clarity, all epochs except 6832~d are multiplied by arbitrary constants.}
\label{fig:mirBB}
\end{figure}

We have performed blackbody fits to the continuum flux densities in the MIR \citep{arendt20} and NIR to investigate an additional component that may be present in the NIR -- assuming only one dominant continuum source. This is likely not true, as in addition to blackbody emission, thermal bremsstrahlung and nonthermal emission can contribute. We fit two blackbody components, the cooler of which is fixed at 525~K \citep{arendt16}. We show these fits and their temperature uncertainties in \autoref{fig:mirBB}. We do not include the $J$-band fluxes, as it is clear from the figure that one blackbody is not sufficient to fit all NIR bands. We use both NACO and SINFONI epochs, but include a correction factor in the latter as described in Sect.~\ref{sec:phot_cont}. Finally, we introduce a 10 per cent uncertainty in the MIR fluxes, which otherwise would be weighted much more heavily than the NIR fluxes in the fit. The temperature of the hotter blackbody component in the resulting fits ranges between 1800 and 2800~K. The individual photometric points show some scatter (see e.g. \autoref{fig:er_lc}), and some epochs (e.g. around 9000~d) are affected by a small dip in $Ks$-band fluxes (possibly due to slight color variation in Star 2), resulting in the highest blackbody fit temperatures. The rest cluster at 1800--2400~K, close to the evaporation temperature of graphite grains \citep[][]{gd89}.

Small-grain dust that is being destroyed over time by photoevaporation could be evident in the NIR data through a blackbody component at a relatively constant (and high) temperature. Grain destruction within the ER could also increase the relative contribution of the outer emission. Despite the expanding diffuse component, most of the NIR continuum is coming from the hotspots, i.e. dense ER clumps, and expanding slower than the MIR continuum emission. Both the warm and hypothetical hot dust would likely have been formed in the same mass loss episode(s) -- however, differences in their location could arise after the SN as graphite grains could survive in hotter locations, and the collisional heating of the dust becomes more efficient in the high-density clumps \citep[e.g.][]{matsuura22}. This could result in a higher relative prevalence of hot graphite dust in the clumps compared to cooler silicate dust.

Assuming that the NIR is dominated by dust, we can estimate the dust mass similarly to \citet{fox10} using their Eq. 2, from which we obtain $M_d = F_\nu D^2 / (B_\nu \kappa)$, where $M_d$ is the dust mass, $F_\nu$ the continuum flux density, $B_\nu$ the fitted hot blackbody component and $\kappa$ a mass absorption coefficient, a function of the grain size and the frequency. With a grain size of 0.1~$\mu$m (below $\sim$0.1~$\mu$m the size has a minimal effect on $\kappa$) we estimate that a hot dust mass on the order of only $\sim$1--4 $\times 10^{-12}~\mathrm{M}_\odot$ could account for the entire $HKs$ continuum, depending on the epoch -- this range corresponds to the scatter in blackbody temperatures. For comparison, \citet{arendt16} estimated the stronger 525~K component to originate from roughly 0.5 per cent of the total dust mass in the ER, which, assuming a total of $10^{-5}~\mathrm{M}_\odot$ at 8000~d \citep{matsuura19}, amounts to $\sim5\times10^{-8}~\mathrm{M}_\odot$ -- we obtain a value of $\sim2\times10^{-8}~\mathrm{M}_\odot$ from the 525~K graphite component in our blackbody fit around the same epoch. \citet{matsuura22} obtained a similar estimate of $\sim10^{-8}~\mathrm{M}_\odot$. A low mass of graphite dust is not surprising based on the low carbon abundance in the CNO-processed ER \citep{lf96}. Since the 525~K component is both stronger and cooler than the NIR continuum (therefore having a much larger emitting area) and has a lower $\kappa$ \citep{fox10}, the $\sim$2000~K dust mass is orders of magnitude lower still. 

However, the $J$-band continuum flux is clearly affected by another component as well; this component can also extend into the $H$ band. The component is already visible in 2006, when the F110W/$J$ ratio is lower than at later times, making it unlikely to be simply the He~{\sc{i}} 1.083~$\mu$m line at the edge of the $J$ filter. The temperature fits above assume that there is only one continuum component in $H$ and $Ks$ and no contribution from e.g. blended weak lines, thermal bremsstrahlung from gas either at the ER or the reverse shock, or a broad-line component from the latter. The blackbody fit is affected in reality by these additional components. The $H$ band is more densely filled with lines than the $Ks$ band, which makes it more difficult to disentangle broad reverse-shock lines from the continuum; thus the effect of reverse-shock lines would be expected to be higher in $H$, which can increase the apparent blackbody temperature. It is likely that the continuum includes a reverse-shock + weak-line component as well. The reverse shock flux in optical lines is a nonnegligible fraction of the total line flux \citep{fransson13,fransson15}. Thermal bremsstrahlung from gas at $\gtrsim$10000~K, typical for the dense ER clumps \citep[e.g.][]{orlando19}, on the other hand, could have a flat spectrum across the $HKs$ bands, which would also distort the temperature from a blackbody fit by increasing the relative contribution in the weaker band (and thus usually the temperature, as this band tends to be $H$). This component may contribute significantly to the total NIR flux, albeit less so in $J$. As a result, the dust mass we obtain is an upper limit only.

Synchrotron emission from the forward or reverse shock may contribute as well. It plays an important role in the radio and, to a lesser extent, FIR \citep[e.g.][]{cendes18,cigan19}. Its extension into NIR (which would be strongest in $Ks$) is a candidate for a relatively fast-expanding diffuse continuum component. Extrapolating the synchrotron spectrum from ALMA frequencies to the NIR with a $\nu^{-0.7}$ power law \citep{cigan19}, one obtains a flux density of $\sim$0.1 mJy (strongest in the $Ks$ band: $\sim$0.13 mJy) at 9280~d, around the peak of the NIR light curve. This is roughly a third of the continuum flux in the NIR at that epoch. The radio fluxes have increased until at least 11000~d (2017), implying that the synchrotron component at the post-peak epochs can constitute even more than a third of the NIR continuum flux. The effect of a synchrotron contribution would be largest in $Ks$, decreasing the apparent blackbody temperature (i.e. possibly offsetting the effects of bremsstrahlung and the weak/broad lines on it). This, however, assumes no break frequency in the spectrum between 1 and $\sim$800~$\mu$m. The ER is not visible in the FIR images of \citet{indebetouw14}, but the upper limits shown in \citet{cigan19} do not constrain whether such a break is present. 

\citet{lundqvist20} used an observed break frequency and the age of the SN remnant SNR 0540-69.3 to estimate its magnetic field; we can similarly estimate that a break frequency $\nu_c$ requires a magnetic field $B = (\tau/6\times10^{11}~\mathrm{s})^{-2/3} \nu_c^{-1/3}$ \citep{pacholczyk70}, where $\tau$ is the lifetime of the synchrotron-emitting electrons. With $\tau\approx15$~yr (at 9280~d, since about the beginning of the ER-ejecta interaction) and $\nu_c > 3\times10^{14}$~Hz, corresponding to 1~$\mu$m, we obtain $B \lesssim 1.8$~mG. \citet{berezhko11} argued that $\gtrsim10$~mG was required to produce the observed radio and X-ray spectrum at 20~yr. This would translate into $\nu_c \sim 6\times10^{11}$~Hz \citep[$\sim500 \mu$m, roughly the lowest value allowed by][]{cigan19} and an insignificant synchrotron component in the total NIR emission. However, \citet{petruk22}, based on the \citet{orlando19} models, place the magnetic field strength much lower than 1~mG in the hotspots and beyond, allowing the synchrotron spectrum to continue unbroken through the NIR. Even in the \citet{berezhko11} scenario, regions of low magnetic field may still contribute, and we note that the continuum ``ring" outside the hotspots clearly does not dominate the total continuum emission either.

%--------------------------------------------------------------------
\section{Conclusions}
\label{sec:concl}

We have examined the morphology of the equatorial ring (ER) of the remnant of SN~1987A in the near-infrared (NIR), its evolution over time, and differences between the NIR and other wavelengths. We have also attempted to isolate the continuum and line emission from the ER and examined differences in morphology between them. Our conclusions are as follows.

   \begin{itemize}
      \item The evolution of the NIR morphology broadly follows that seen in the optical, with a similar light curve, expansion speed as measured from hotspots ($\sim700$ km~s$^{-1}$), and a decreasing east-west brightness ratio. However, the NIR emission at later epochs is stronger outside the ring defined by the hotspots than in F625W (but still weaker than the hotspots), hinting at a faster-expanding diffuse component.
      \item Although the line and continuum emission follow similar light curves, the continuum emission is similarly stronger outside the hotspots than line emission in the later epochs. This indicates that the continuum is responsible for the outer emission not seen in the optical.
      \item Continuum emission constitutes a majority of the total flux in the NIR in all bands and at all epochs, but not all bands can be simultaneously fitted using a blackbody function. If the $HKs$-band continuum is dominated by dust, its temperature is in the vicinity of 2000~K from 2005 to 2017. The location of this hypothetical hot dust would also be different from the warm dust dominating in the mid-infrared, which expands much faster and is more diffuse in morphology. This dust does not dominate in the $J$ band, however, and thermal bremsstrahlung is also expected to contribute; thus the mass of any hot dust is not more than a few $\times10^{-12}$~M$_\odot$ -- orders of magnitude below the warm dust mass. 
      \item Synchrotron emission can contribute significantly to the NIR continuum in regions where the magnetic field is below $\sim$2~mG, especially in the $Ks$ band. It cannot dominate the full NIR continuum emission in the ER, which mostly originates in the hotspots, but can be the source of the outer continuum emission.
   \end{itemize}
   
With the recent launch of the \emph{JWST} with its superior spatial resolution and sensitivity, the community is now able to study the ER and other regions in the remnant of SN~1987A in greater detail. These observations will no doubt cast more light on the different emission components in the NIR and allow for better constraints on their origins.

\begin{acknowledgements}

We thank the anonymous referee for their suggestions. We thank John Danziger, Jason Spyromilio, Jesper Sollerman, Anders Jerkstrand, Roger Chevalier and Rubina Kotak for contributing to the NACO observations and for helpful discussion. This work was supported by the Swedish National Space Agency and the Swedish Research Council. S. M. was funded by the Academy of Finland project 350458.

This work is partially based on observations made with the NASA/ESA \textit{Hubble Space Telescope} (programmes GO 7434, 7821, 9248, 10549, 10867, 11653, 12241, 13181, 13810 and 14753), obtained through the data archive at the Space Telescope Science Institute (STScI). STScI is operated by the Association of Universities for Research in Astronomy, Inc. under NASA contract NAS 5-26555.
      
\end{acknowledgements}

% WARNING
%-------------------------------------------------------------------
% Please note that we have included the references to the file aa.dem in
% order to compile it, but we ask you to:
%
% - use BibTeX with the regular commands:
%   \bibliographystyle{aa} % style aa.bst
%   \bibliography{Yourfile} % your references Yourfile.bib
%
% - join the .bib files when you upload your source files
%-------------------------------------------------------------------
\bibliographystyle{aa}
\bibliography{kangas_87A_ER_NIR}
\clearpage

\end{document}